\documentclass{aa}  

\usepackage{graphicx}
\usepackage{txfonts}
\usepackage[binary-units=true]{siunitx}
\DeclareSIUnit\pixel{px}
\begin{document}
   \title{Initial radiometric calibration of the High-Resolution EUV Imager ($\textrm{HRI}_\textrm{EUV}$) of the Extreme Ultraviolet Imager (EUI) instrument onboard Solar Orbiter}

   \titlerunning{Initial radiometric calibration of $\textrm{HRI}_\textrm{EUV}$}

   \author{
        S. Gissot\inst{\ref{i:rob}} \fnmsep\thanks{Corresponding author: Samuel Gissot \email{samuel.gissot@oma.be}}
        \and
        F. Auch\`ere\inst{\ref{i:ias}}
        \and
        D. Berghmans\inst{\ref{i:rob}}
        \and
        B. Giordanengo\inst{\ref{i:rob}} %TODO now at...
        \and
        A. BenMoussa\inst{\ref{i:eu_iter}}
        \and
        J. Rebellato\inst{\ref{i:iogs}}
        \and
        L. Harra\inst{\ref{i:pmod}}
        \and
        D. Long\inst{\ref{i:mssl}}
        \and
        P. Rochus\inst{\ref{i:csl}}
        \and
        U. Sch\"uhle\inst{\ref{i:mps}}
        \and
        R. Aznar Cuadrado\inst{\ref{i:mps}}
        \and
        F. Delmotte\inst{\ref{i:iogs}}
        \and
        C. Dumesnil\inst{\ref{i:ias}}
        \and
         A. Gottwald\inst{\ref{i:ptb}}
        \and
        J.-P. Halain\inst{\ref{i:csl}}
        \and 
        K. Heerlein\inst{\ref{i:mps}}
        \and
        M.-L. Hellin\inst{\ref{i:csl}}
        \and
        A. Hermans\inst{\ref{i:csl}}
        \and
        L. Jacques\inst{\ref{i:csl}}
        \and 
        E. Kraaikamp\inst{\ref{i:rob}}
        \and
        R. Mercier\inst{\ref{i:iogs}}
        \and
        P. Rochus\inst{\ref{i:csl}}
        \and 
        P. J. Smith\inst{\ref{i:mssl}}
        \and 
        L. Teriaca\inst{\ref{i:mps}}
        \and
        C. Verbeeck\inst{\ref{i:rob}}
        }
        \institute{
            Royal Observatory of Belgium, Ringlaan -3- Av. Circulaire, 1180 Brussels, Belgium\label{i:rob}
            \and
            Institut d'Astrophysique Spatiale, CNRS, Univ. Paris-Sud, Universit\'e Paris-Saclay, B\^at.\ 121, 91405 Orsay, France\label{i:ias}
            \and
            Centre Spatial de Li\`ege, Universit\'e de Li\`ege, Av. du Pr\'e-Aily B29, 4031 Angleur, Belgium\label{i:csl}
            \and
            Max Planck Institute for Solar System Research, Justus-von-Liebig-Weg 3, 37077 G\"ottingen, Germany\label{i:mps}
            \and 
             European Commission, DG Energy, Directorate D: Nuclear Energy, Safety and ITER, 24 rue Demot, B-1049 Brussels, Belgium \label{i:eu_iter}
            \and
            UCL-Mullard Space Science Laboratory, Holmbury St.\ Mary, Dorking, Surrey, RH5 6NT, UK\label{i:mssl}
            \and
            Physikalisch-Technische Bundesanstalt (PTB), Abbestr. 2-12, 10587 Berlin, Germany\label{i:ptb}
            \and
            Physikalisch-Meteorologisches Observatorium Davos, World Radiation Center, 7260, Davos Dorf, Switzerland\label{i:pmod}
            \and
            Laboratoire Charles Fabry, Institut d'Optique Graduate School, Universit\'e Paris-Saclay, 91127 Palaiseau Cedex, France\label{i:iogs}}

   \date{Received ; accepted }

  \abstract
   {The $\textrm{HRI}_\textrm{EUV}$ telescope was calibrated on ground at the Physikalisch-Technische Bundesanstalt (PTB), Germany's national metrology institute, using the Metrology Light Source (MLS) synchrotron in April 2017 during the calibration campaign of the Extreme Ultraviolet Imager (EUI) instrument onboard the Solar Orbiter mission.
   }
   {We use the pre-flight end-to-end calibration and component-level (mirror multilayer coatings, filters, detector) characterization results to establish the beginning-of-life performance of the $\textrm{HRI}_\textrm{EUV}$ telescope which shall serve as a reference for radiometric analysis and monitoring of the telescope in-flight degradation. }
   {Calibration activities  at component level and end-to-end calibration of the instrument were performed at PTB/MLS synchrotron light source (Berlin, Germany) and the SOLEIL synchrotron.
   %TODO check the SOLEIL ref.
   Each component optical property is measured and compared to its semi-empirical model. This pre-flight characterization is used to estimate the parameters of the semi-empirical models. The end-to-end response is measured and validated by comparison with calibration measurements, as well as with its main design performance requirements. }
   {The telescope spectral response semi-empirical model is validated by the pre-flight end-to-end ground calibration of the instrument. It is found that $\textrm{HRI}_\textrm{EUV}$ is a highly efficient solar EUV telescope with a peak efficiency superior to 1 e$^-$.ph$^{-1}$), low detector noise ($\approx$ 3 e- rms),  low dark current at operating temperature, and a pixel saturation above 120 ke- in low-gain or combined image mode. The ground calibration also confirms  a well-modeled spectral selectivity and rejection, and  low stray light. The EUI instrument achieves state-of-the-art performance in terms of signal-to-noise and  image spatial resolution. }
   {}


 \maketitle

\section{Introduction}
The EUI instrument is a suite of three telescopes, consisting of  the Full-Sun Imager (FSI),  a large field-of-view Extreme Ultraviolet (EUV) telescope, the high-resolution imager Lyman-$\alpha$ telescope $\textrm{HRI}_\textrm{LYA}$, and the HRI Extreme Ultraviolet (EUV) telescope  ($\textrm{HRI}_\textrm{EUV}$) \citep{Rochus2020} onboard the Solar Orbiter spacecraft that was launched  on February 11, 2020.

$\textrm{HRI}_\textrm{EUV}$ is a high-resolution telescope imaging the solar corona with a design peak wavelength at 17.4 nm, where the Fe IX/X spectral lines dominate the solar signal which are 
%\david{dominate x2} by solar plasma collisional excitation. 

The $\textrm{HRI}_\textrm{EUV}$ channel \citep{Halain2014, Halain2015} is an off-axis Cassegrain telescope optimized in length and width to fulfill stringent mass and power restrictions imposed by the specific Solar Orbiter constraints. Thanks to the Solar Orbiter orbits around the Sun,  it is able acquire long sequences of EUV images of the solar corona at an image resolution never achieved before. Furthermore in the later phase of the mission, the out-of-ecliptic view point  will enable observations high-resolution observations of the polar regions and active latitudes from above. A recent overview of the observational capabilities of $\textrm{HRI}_\textrm{EUV}$  is presented in \citep{Berghmans2023}. 

The goal of the end-to-end calibration that occurred in April 2017 is to validate the design performance requirements before flight. The results of the pre-flight calibration is also a unique opportunity to perform the initial calibration which shall be used as a reference for the study of the in-flight telescope performance.

The characterization and the modelling of the instrument radiometric response relies on measurements at optical component level (mirror multilayer coatings, filters, detector) from past flight-model component characterization campaigns and on the ground calibration. 

This is achieved by performing ground measurements and defining semi-empirical models for each optical component. 
The end-to-end response is then modelled by combining component-level model in a  spectral response semi-empirical model, and shall be used to establish the beginning-of-life telescope detailed performance.
In this work early in-flight data acquired during instrument commissioning phase are also used as complementary measurements to characterize the $\textrm{HRI}_\textrm{EUV}$ image sensor and complete the pre-flight measurements.

Among the comparable instruments observing at similar wavelength, SoHO/EIT \citep{Delaboudiniere1995, Dere2000} observing at L1 Lagrangian point had a plate scale of  \SI{2.63}{\arcsecond\per\pixel}.   
The TRACE instrument \citep{Handy1999} has an on-axis Cassegrain design with a plate scale of \SI{0.5}{\arcsecond\per\pixel}.
The STEREO/SECCHI EUVI instruments \citep{Wuelser2007} acquires full sun images with a plate scale of \SI{1.6}{\arcsecond} and was the first EUV imager to image away from the Sun-Earth line.
The AIA instrument \citep{Lemen2011}  onboard SDO observes at  17.1 nm and has a spatial resolution of  \SI{0.6}{\arcsecond\per\pixel}.
SDO/AIA has achieved an overall accuracy of pre-flight calibration of the order of 25\% \citep{Boerner2011} relying on a sub-system characterization and no end-to-end calibration. 
The Hi-C 2.1 instrument \citep{Rachmeler2019}, an evolution of the Hi-C telescope \citep{Kobayashi2014},  has observed the EUV corona at similar wavelength (17.2 nm peak wavelength), with 0.1 arcsec/pixel spatial resolution for Hi-C 1 and 0.13 arcsec/pixel, represents an improvement of 0.6 to 0.13 arcsec/pixel, equivalent to 200 km two-pixel resolution using a design and components similar to AIA.
SWAP  was the first off-axis EUV imager with a complementary metal oxide semiconductor (CMOS)  Active Pixel Sensor  (APS) image detector that  provides images of the solar corona at about 17.4 nm at a cadence of 1 image per 1-2 minutes, and field of view (FOV) of 54 arcmin \citep{Seaton2012, Halain2012}.
% SPIRIT

On-ground calibration of an EUV telescope is a challenging task since it requires an EUV calibration light source such as an electron-storage ring synchrotron light source and the technical difficulties include the instrument accommodation, high vacuum, cooling,  image centering and beam localization. Using the synchrotron radiation of the MLS storage ring \citep{Gottwald2010} dedicated to EUV metrology, designed and built to meet dedicated demands for metrology  synchrotron absolute calibration  of $\textrm{HRI}_\textrm{EUV}$ is possible  in the EUV  wavelength range  (1 nm-100 nm) at an accuracy of 5 \% or below. 

Strict cleanliness requirements for the EUI instrument were fulfilled to avoid any contamination that would affect the scientific performance of the instrument \citep{Rochus2020, Benmoussa2013a}.

After an overview of the $\textrm{HRI}_\textrm{EUV}$ telescope (Section \ref{sec:telescope_overview}), the results of component-level characterization and modelling is presented (Section 
\ref{sec:component_characterization}). The validation of this semi-empirical model are compared to pre-flight ground end-to-end calibration of the telescope (Section \ref{sec:ground_calibration}) using EUV synchrotron light beam completed with  first light and in-flight calibration data acquired during commissioning. The instrument spectral response is discussed in Section \ref{sec:spectral_response}, followed by conclusions in Section \ref{sec:conclusion}.

\section{Telescope overview}
\label{sec:telescope_overview}
$\textrm{HRI}_\textrm{EUV}$ is a high-resolution telescope imaging the solar corona with a design peak wavelength of 17.4 nm. It is based on an off-axis Cassegrain telescope optimized in length and width \citep{Halain2014, Halain2015}. 

Among the challenging $\textrm{HRI}_\textrm{EUV}$ design requirements, the  spectral purity shall be greater than 90\% for the Fe IX (17.11 nm) and Fe X (17.45 nm and 17.72 nm) ionization states when observing quiet Sun and active region. 

Table \ref{tab:t1} summarizes the main performance requirements derived from the scientific specification from the early phase of the telescope design.

\begin{table}
\caption{\label{tab:t1} $\textrm{HRI}_\textrm{EUV}$ main performance requirements.}
\centering
\begin{tabular}{lc}
\hline\hline
Requirement & Value (target)\\
\hline
Peak wavelength & 17.4 nm \\
Total FOV & 1000 arcsec square \\
Plate scale  & $\simeq$ 0.5 arcsec/pixel \\
Angular resolution (2-pixel) &  1 arcsec \\
Spectral purity & 90\%  \\

\hline
\end{tabular}
\end{table}

\subsection{Optical design}
\label{sec:optical_design}

\begin{figure*}
   \includegraphics[width=\linewidth]{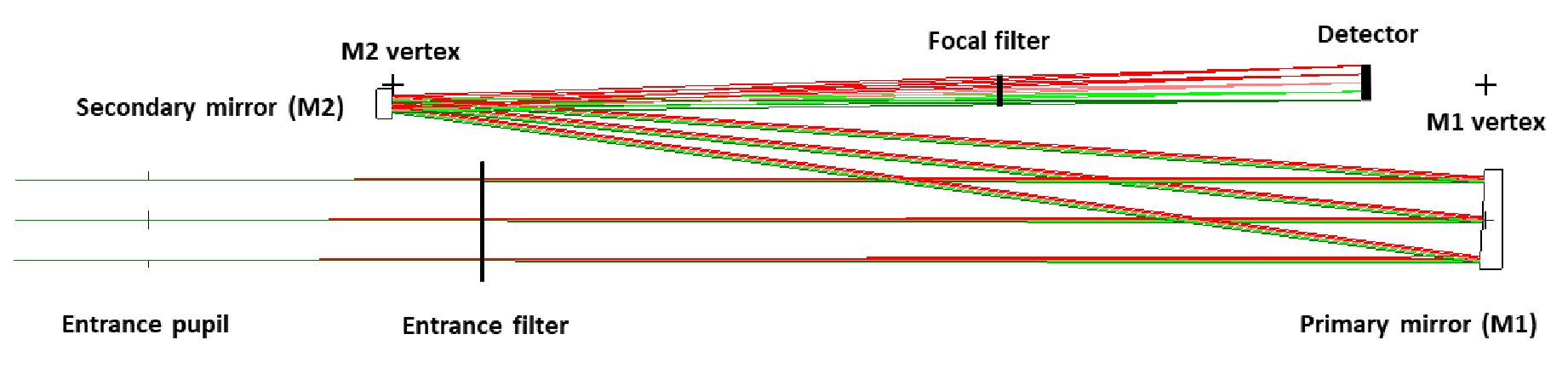}
  \caption{\ $\textrm{HRI}_\textrm{EUV}$ optical design. The $\textrm{HRI}_\textrm{EUV}$ channel \citep{Halain2015} is an off-axis Cassegrain telescope optimised in length and width. The entrance pupils are located at the front section of dedicated entrance baffes and have a diameter of 47.4 mm.}

  \label{fig:hrieuv_EUV_optical_design}
\end{figure*}

The off-axis Cassegrain  optical design of the $\textrm{HRI}_\textrm{EUV}$ telescope consists of a combination of a primary parabolic mirror and a  hyperbolic convex mirror (see Fig. \ref{fig:hrieuv_EUV_optical_design}) \citep{Halain2015}.

The optical structure is made out of a Carbon Fibre Reinforced Polymer sandwich panel with an aluminium honeycomb core to meet stiffness and thermal stability requirements while achieving a low mass. 

The solar light enters the instrument through the heat shield feedthrough designed to prevent light reflection inside the entrance pupils and to limit the straylight while avoiding additional heat load.
The reflective entrance baffle reject more than 60\% of the heat coming through the entrance pupil.

The entrance pupil (Figure \ref{fig:hrieuv_entrance_filter}) is located at the front section of the entrance baffle and has a diameter of 47.4 mm.

The EUV reflective coatings of the mirrors are specific multilayers optimized to provide the optimal spectral passbands. Their design takes into account the angle of incidence on the mirrors.

A filter wheel, comprising two EUV filters (nominal and redundant) and  open positions, is located at the output pupil in front of the detector. It is used a redundancy to entrance filter, but it provides an occulting (or ``blocking'') position. These filters also have an effect on the spectral purity of the telescope spectral response.

The front and rear EUV filter are of primary importance for the instrument to avoid contamination with visible light that can be $10^8$ times more intense than the EUV flux.

The design of the $\textrm{HRI}_\textrm{EUV}$ telescope was severely constrained by the Solar Orbiter spacecraft design constraints. 
The constraints on dimensions include a total length including mechanism and FPA inferior to  820 mm and a maximum width inferior to 120 mm. Other optical design constraints included an incidence angle onto mirror coating inferior to 6 arcdeg. In order to achieve a sufficient signal-to-noise ratio (SNR), the entrance pupil diameter was chosen to be 47.4 nm.

The EUI instrument is an internally mounted unit.  The optical structure is a thermally insulated unit to remains in an acceptable temperature range in hot and cold flight environment, with low conductive link to the platform (low conductance mounts) and radiative insulation (MLI). 
The flux incoming through the aperture is mostly rejected by the entrance baffles. The remaining absorbed flux by the entrance baffle is dissipated through the hot element interface and radiatively to the S/C cavity. 
The detector is cooled at a temperature lower than -40 \textcelsius (target -60 \textcelsius) with a stability of $\pm$ 5 \textcelsius.

The low allocated telemetry  is 20 kbits/s on average which implies that lossy compression must be implemented in order to optimize the telemetry allocation. As a consequence the image must be reconstructed onboard in order to perform high-quality onboard lossy compression.

The telescope components and the status of their measurements are detailed in Table \ref{tab:hrieuv_components}. 
Measurements were made on similar reference components rather than on the flight model as detailed in \cite{Auchere2013}. 

\begin{table}
\center
\caption{Overview of $\textrm{HRI}_\textrm{EUV}$ subsystems (filters and mirror) with their references and spectral response measurement ranges. The M1 mirror refers to MP15041-f40\_P2\_V  and M2 mirror refers to MP15057-s12\_S1\_V2.}
\begin{tabular}{l|c|c} 
\hline
On-board  & Measured & Measurement \\ 
component &  reference & range\\ 
\hline
Entrance filter Al & Al $\textrm{HRI}_\textrm{EUV}$ - 152.7 nm & 5.55 to \SI{40}{\nano\metre}\\ 
Mirror M1 & & \\ 
Mirror M2 & &  \\ 
Al n$^{\circ}$ SN07\_V3 & 149.2 nm & \\
Al n$^{\circ}$ SN10\_V3 & 149.2 nm & \\
\end{tabular}
\label{tab:hrieuv_components}
\end{table}

The entrance pupil is located at the front section of the entrance baffles and has a diameter of 47.4 mm for the $\textrm{HRI}_\textrm{EUV}$. For thermal and mechanical reasons, the length of the entrance baffle  is $\approx$  100 mm long with an aperture diameter equal to the entrance pupil diameter (47.4 mm for HRIEUV) and a cone angle of 2.2 arcdeg. This design presents a very small distortion, lower than 0.1 \% that is  not symmetric  due to the off-axis design.

\subsection{Entrance filter}

\begin{figure}
 \centering
 \includegraphics[width=7cm]{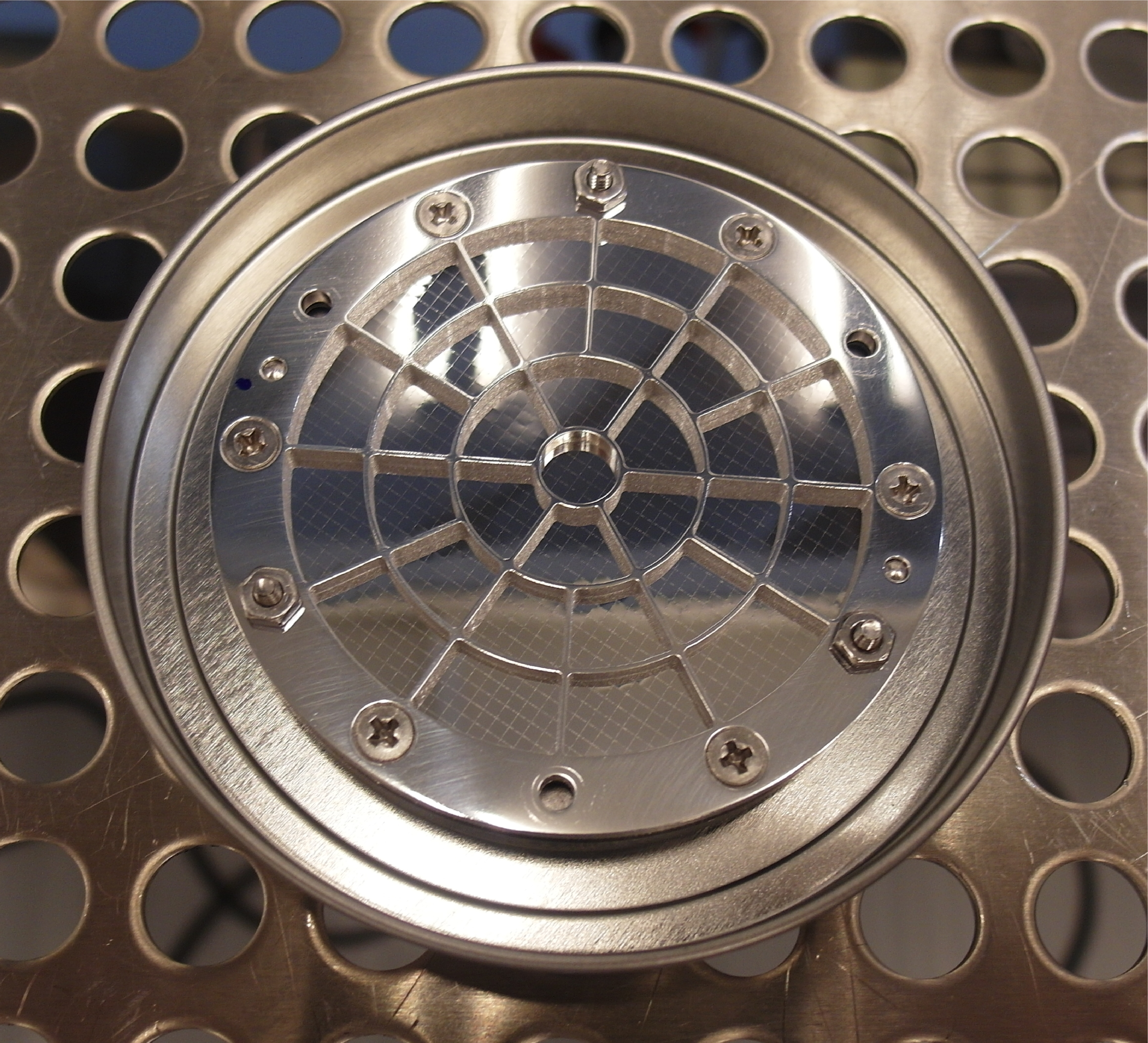}
  \vspace{10pt}
  \caption{\label{fig:entrance_filter} $\textrm{HRI}_\textrm{EUV}$ entrance filter.}

  \label{fig:hrieuv_entrance_filter}
  
\end{figure}

The  $\textrm{HRI}_\textrm{EUV}$ entrance filter is located at the end of the front baffle and is manufactured by Luxel. It consists an aluminium foil filter inserted between the entrance aperture (entrance pupil) and the primary mirror to provide protection against excessive heat input on the mirror and efficient rejection of the visible light  with a visible light transmission of $9 \times 10^{10}$ according to Luxel. It consists of an 150 nm Al thin-foil film supported by custom 20 lpi Ni patterned mesh, mounted (epoxy free) on a webbed Aluminum frame with a 50mm inner diameter (see Fig. \ref{fig:entrance_filter}). 

The rib structure divides the filter area into sectors to improve thermal conductivity and mechanical resistance (see Figure \ref{fig:hrieuv_entrance_filter}) to protect agains excessive heat input on the mirror and rejects the visible light. 

\subsection{Mirror coatings}
The  $\textrm{HRI}_\textrm{EUV}$  mirrors use a periodic multilayer coating with peak reflectivity at 17.4 nm. Provided by the Laboratoire Charles Fabry, Institut d'Optique (LCF/IO), they  are based on periodic Al/Mo/SiC multi-layer coating with first order centred at 17.4 nm optimized for high efficiency and desired spectral pass-bands capped with a  3 nm SiC  layer for temporal stability \citep{Delmotte2013}.
The main characteristics of these mirrors are summarized in Table \ref{tab:perf_summary}.

%TODO Add coatings description

\begin{table}
\caption{\label{tab:t3} $\textrm{HRI}_\textrm{EUV}$ mirror where R is the mirror radius of curvature at apex and K is the conic constant.}
\centering
\begin{tabular}{lc}
\hline\hline
Primary diameter & 66 mm  (54 mm useful diameter)\\
& 80 mm off-axis\\
R & 1518.067 mm (concave)\\
K & -1\\
\hline\hline
Secondary diameter & 25 mm  (12 mm useful diameter)\\
&11.44 mm off-axis\\
R & 256.774 mm (convex)\\
K & -2.0401375\\
\hline
\end{tabular}

\end{table}

The $\textrm{HRI}_\textrm{EUV}$ point spread function (PSF) is difficult to measure because during calibration, the synchrotron EUV beams has a non-negligible divergence.
By design (RMS spot diameter), the PSF width is smaller than the pixel size (10 $\mu$m) over the detector field of view.
The PSF core originates from the optical system aberrations and mirror micro-roughness. The entrance filter Ni mesh grid also contributes to the PSF function.

%TODO include PSF models image

The stray light level in the EUV (resp. visible) wavelengths is specified to be 1e-4 times (resp. 1e-8) times lower than the flux of the nominal image in the same pixel.

\subsubsection{Focal plane filters}

A filter wheel with four filter slots is located at the focal plane (exit pupil),
including two redundant filters of similar mesh grid as the entrance filter and two open positions. 
The focal plane filters are Al foils similar to the EUV entrance filter. 
The filter wheel  has three operating modes, either in  filter position (nominal and redundant filters), or blocking position (both filters in intermediate position) or opened position.

These filters maybe particularly useful in case a pinhole appears on the entrance filter, and thus mitigate the effect of a pinhole light leakage.

In intermediate position, the filter wheel can be used as a blocking position to protect the detector or make calibration measurements without having to close the telescope internal door. 

The transmission of the Aluminum filter was measured at PTB/MLS synchrotron light source. 

\subsection{CMOS APS image sensor}

\begin{figure}
   \includegraphics[width=\linewidth]{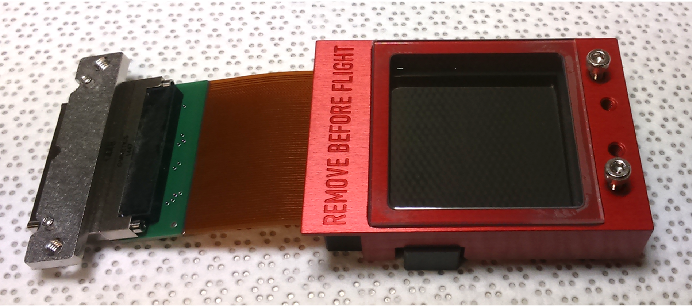}
\vspace{10pt}
  \caption{$\textrm{HRI}_\textrm{EUV}$ CMOS APS detector.}

  \label{fig:hrieuv_cmos_aps}
\end{figure}

The $\textrm{HRI}_\textrm{EUV}$ detector is the APSOLUTE2 image sensor,  a back-side illuminated image sensor made in 0.18 micrometer CMOS process  based on APSOLUTE1 detector \citep{Benmoussa2013} with 10 micrometer pixel pitch. 
APSOLUTE2 has a 3Kx3K image format, optimized EUV sensitivity and low noise (Fig. \ref{fig:hrieuv_cmos_aps}). The $\textrm{HRI}_\textrm{EUV}$ detector is manufactured by CMOSIS (now AMS), a Belgian company.

%The spare detector used for $\textrm{HRI}_\textrm{EUV}$ is the D30 (T409754-((01)) = T409754-11 = DDQJB082WPC0 X001Y002).

The device is made with two stitching blocks of 3072 x 1536 pixels based on a dual-gain pixel architecture. 
Each pixel structure is based on a 4 transistors (4T) photodiode pixel design and their in-pixel transistors differ in shape and size. 
The EUV sensitivity is achieved with backside illumination on a Silicon-on-Insulator (SOI) material based solution manufactured by SOITEC (France). 
The detector has a non-destructive readout with correlative double sampling (CDS) that removes most of the reset noise.
In order to reach ambitious goals for both the read noise and the dynamic range, a dual-gain pixel design was developed: each pixel outputs both a high-gain signal, with low read noise and a low saturation level, and a low gain signal, with a high saturation level but larger read-noise.

The pixel design is based on Deep Trench Isolation technology first used in CMOS image sensors by IMEC (2007), ST Microelectronics and Samsung. Although it has the advantage of reducing optical and electrical cross-talk, it may also increase the dark current.

%\david{Do we need a reference for the previous two sentences?}

The CMOS-APS detectors have an electronic rolling shutter that requires no mechanism and allow for an integration time ranging from 1 ms to several minutes.
The radiation tolerance of these detectors was demonstrated in \citep{Benmoussa2013}. 
Its power consumption remains below 1 Watt.

The APS sensors and the proximity electronics are located behind a cold cup  in front of the detector which acts as a field stop and limits the light within $2348 \times 2340$ pixels in the detector plane. 

The EUI is a compact instrument based on a passive  thermal control designed to maintain the detector temperature below -30 \textcelsius (-50 \textcelsius reached during flight) slightly warmer than the surrounding cold cup  to avoid condensation on the detector, which would degrade the instrument response.

The instrument contains one nominal heater per detector to enable in-flight bake-out and annealing. The detector has an internal temperature sensor that has been calibrated on-ground. During bake-out  campaigns detector temperature is controlled by the CEB following a configurable duty-cycle
to ensure the temperature of the detectors does not exceed the
maximum non-operating temperature.

\begin{table}
\caption{\label{t_detector} Detector requirement specification.}
\centering
\begin{tabular}{lcc}
Parameter \vline \vline &  &  \\
\hline\hline
Sensitivity (EUV QE) & >50\%\\
Image size & > 2Kx2K\\
Full Well Capacitance & > 80 ke\\
Dark current & < 1 e.s$^{-1}$ \\
Read noise & < 5 e rms \\
Max cadence & 1s\\
Power & $< 1$ W
\end{tabular}

\end{table}

% The sensor register settings were chosen to optimize the offset values in order to match onboard computation constraints and enable the onboard image calibration and dual-gain reconstruction.
% The gain factor was left as is while the clipping values were also optimized.
% The detectors have a full well of the order of 120k electrons
% (device dependent), which is the same for both gains (defined by
% saturation of the pinned photodiode), but the high gain is ADC
% clipped. The theoretical ratio in gain between the high and low
% gain channels with default electronic settings is 22.3. In practise,
% this ratio is device dependent and configurable in flight

\subsection{Onboard calibration LED}
\label{sec:calibration_LED}
Facing the sensor from the side of the cold cup,  two onboard calibration LED (main and redundant) are used to monitor the detector performance in-flight. The LEDs are Micropac 62087-305 which emit at a peak wavelength of 470 nm. They are used  for ground tests and in-flight relative calibration by oblique illumination of the detector.

%TODO LED graph

\section{Component characterization}
\label{sec:component_characterization}
In this section, all the pre-flight component-level measurements are reported and analyzed.

\subsection{Spectral response model}
The instrument response $R$ is given by the formula: 

\begin{equation}
\label{eq:spectral_response}
 R(\lambda) = D(\lambda)  T_{FP}(\lambda)  R_{M2}(\lambda) R_{M1}(\lambda) T_{EF}(\lambda) A_{geom}
\end{equation}

where $A_{geom}$ is the entrance aperture, $T_{EF}$ is the entrance filter transmission, and $D$ is the detector response (or efficiency). $T_{FP}$ is the focal plane filter (two redundant filter position on a rotating wheel in front of the sensor. It is equal to $1$ when the filter wheel is in open position (nominal during ground calibration) and $0$ when in ``blocking position''.
In flight, in order to protect the sensor from potential strong EUV light-flux damage and to lower the EUV-induced contamination, the filter wheel is during nominal mode in the Al-filter position.

The overall optical efficiency $E$ is given by formula
\begin{equation}
\label{eq:spectral_response_efficiency}
 E(\lambda) = T_{EF}(\lambda) R_{M1}(\lambda) R_{M2}(\lambda) T_{FP}(\lambda)  D(\lambda)  
\end{equation}

In order to measure the response function, it is necessary to measure the geometric area, the entrance and focal plane  EUV filter transmittances, the two mirror reflectances, and the APS image detector response. 
%TODO  contamination

\subsection{Entrance filter transmittance}

The filter spectral transmittance was measured at MLS/PTB, Germany in the grazing incidence beamline and at the SOLEIL synchrotron (Orsay, France).
%\david{Do this facilities have descriptive papers that can be cited?}
%TODO TBC

\begin{figure}
\centering
 \includegraphics[width=\linewidth]{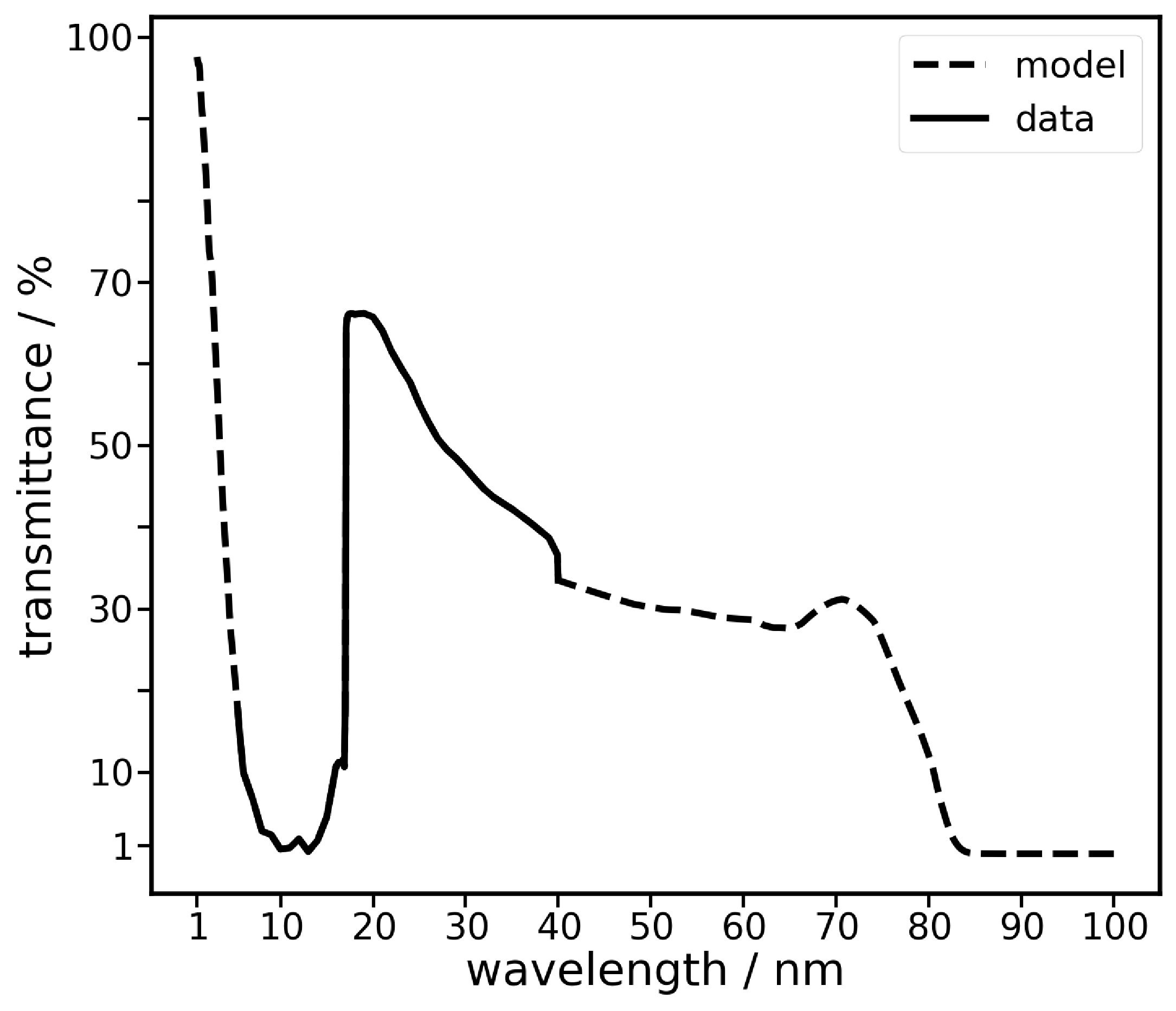}

  \caption{Transmittance of the $\textrm{HRI}_\textrm{EUV}$ entrance filter. The dashed line corresponds to the transmittance measured during the characterization campaign.}

  \label{fig:hrieuv_entrance_filter_transmittance}
\end{figure}

The transmittance of the entrance filter is shown in Fig. \ref{fig:hrieuv_entrance_filter_transmittance}.

\subsection{Mirror multilayer coating reflectance}

Because of its Cassegrain design, $\textrm{HRI}_\textrm{EUV}$ has two multilayer coated mirrors (see Section \ref{sec:optical_design}).

\begin{figure}
\centering
 \includegraphics[width=7cm]{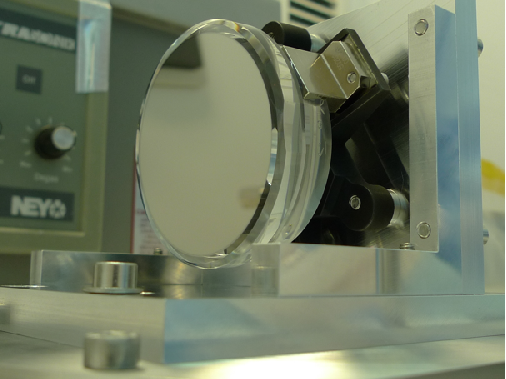}
\vspace{10pt}
  \caption{Picture of the $\textrm{HRI}_\textrm{EUV}$ primary mirror flight model before integration in the EUI instrument.}

  \label{fig:hrieuv_primary_mirror_picture}
\end{figure}

%TODO Multilayer coating X-Y scanning
The M1 mirror is a convex mirror with a useful diameter of 54 mm and that is 80 mm off axis (see Fig. \ref{fig:hrieuv_primary_mirror_picture}).

\begin{figure}
\centering
 \includegraphics[width=7cm]{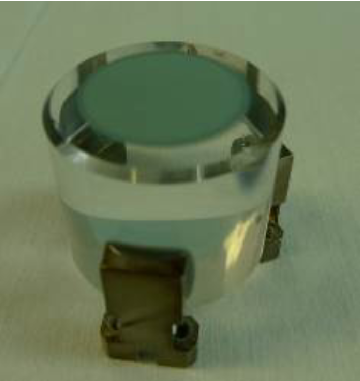}
\vspace{10pt}
  \caption{Picture of the $\textrm{HRI}_\textrm{EUV}$ secondary mirror flight model before integration in the EUI instrument.}

  \label{fig:hrieuv_secondary_mirror_picture}
\end{figure}

\subsubsection{Mirror reflectance}

The mirror reflectance has been modeled and part of its spectral reflectivity was measured at SOLEIL Synchrotron. 
\begin{figure}
\centering
 \includegraphics[width=\linewidth]{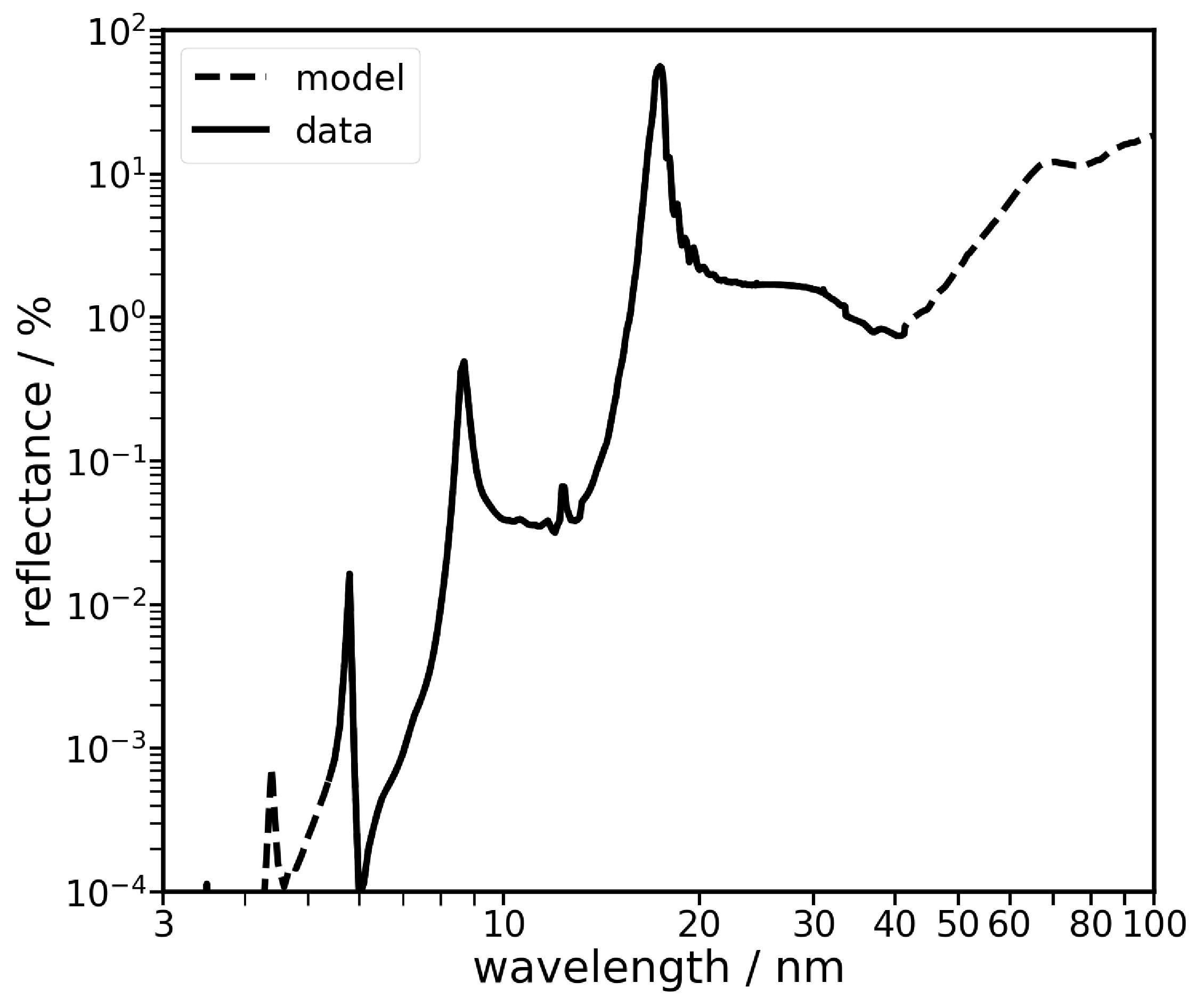}
  \caption{Reflectance of the $\textrm{HRI}_\textrm{EUV}$ primary mirror M1. The  dashed line part of the curve corresponds to the reflectance measured during the characterization campaign.}

  \label{fig:hrieuv_primary_mirror}
\end{figure}

\begin{figure}
\centering
 \includegraphics[width=\linewidth]{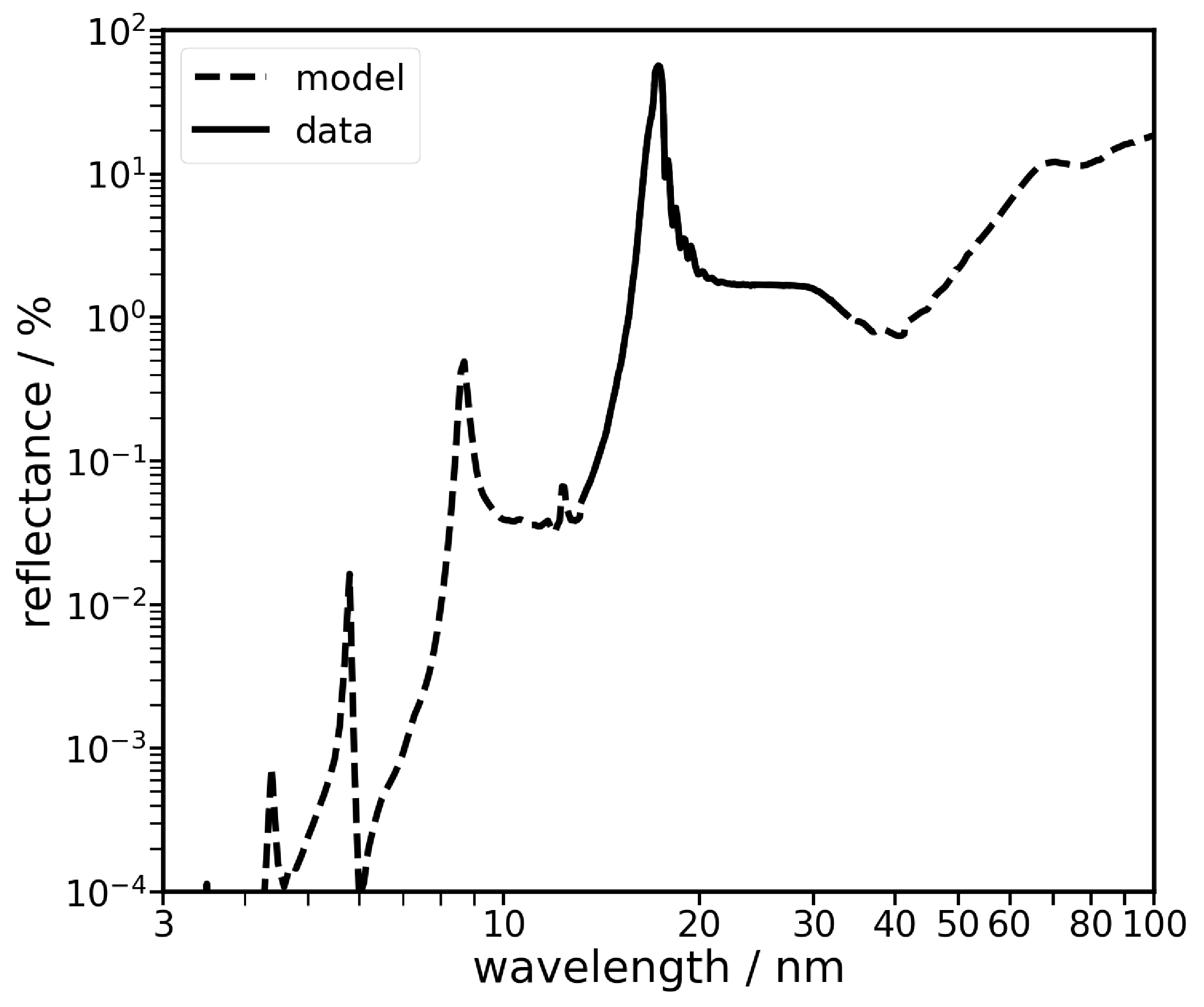}

  \caption{Reflectance of the $\textrm{HRI}_\textrm{EUV}$ secondary mirror M2. The  dashed line part of the curve corresponds to the reflectance measured during the characterization campaign.}

  \label{fig:hrieuv_secondary_mirror}
\end{figure}

The reflectance of the M1 primary mirror (resp. M2) is shown in Figure \ref{fig:hrieuv_primary_mirror} (resp. Figure \ref{fig:hrieuv_secondary_mirror}).

\subsection{Focal plane filters}

The focal filters are mounted on a filter wheel close to the focal plane and in front of the sensor.
There are two  similar Al filters in opposite positions and two open positions (see Figure \ref{fig:hrieuv_focalplane_filters}).

Figure \ref{fig:spectral_response_2} show their influence on the telescope response, in particular on the off-band rejection.

\subsection{APS detector characterization}

The front-end electronics that controls the APS sensor acquires the detector signal with a 14-bit ADC that rounds the signal to 12 bits. 
Due to its multiplexing architecture, the fixed-pattern noise of the detector is dominated by a 4-column repeating pattern. 

%TODO explain and show the 4-col pattern
%TODO Laser annealing
Before calibration, the detectors were ranked and the flight model was chosen based on the results of the calibration campaign that occurred in January 2017. 
During the integration of the instrument, the selected detector D22 was broken and had to be replaced by detector D30.
This detector is has received an improved surface treatment to clean the SiO2 native oxide at the back surface of the detector, on the illuminated side.
Unfortunately no EUV flat-field had been determined for the D30 detector in the January 2017 campaign at IAS.
As a fall-back options it was decided to use the onboard ``blue'' LED, with a peak emission at 470 nm in order to derive the flat-field required by the ground calibration procedure. 
%TODO dual -gain description
For a correct image reconstruction, required for high dynamic range applications, it is required to assess correctly the gain ratio between both gains. It can then be applied onboard to reconstruct the image correctly, before applying any lossy compression.

It should be noted that the ground characterization of the EUI FM detectors was done using the EGSE provided by the manufacturer, and images acquired in continuous mode, with systematic acquisition of both gain images at the same time (from the same exposure).

\begin{figure}
\centering
 \includegraphics[width=\linewidth]{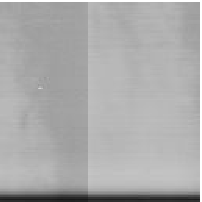}
\vspace{30pt}
  \caption{Average dark signal image at 2-s integration time of the high-gain image acquired in-flight.
  The signal is dominated by a 4-column repeating pattern due to the column gain multiplexing.
  The read noise is low due to averaging and the dark current is negligible at operating temperature (-50 \textcelsius).} 
  The stiching line of the detector, is visible here as a column with an off-point due to the optimized $2K\times2K$ centering. 
  \label{fig:hrieuv_dark_signal}
\end{figure}

The dark signal acquired at 2 second integration time is shown in Figure \ref{fig:hrieuv_dark_signal}, 
with best detector settings optimized in-flight and optimized $2K\times2K$ centering. 

% TODO read noise figures

\subsubsection{Dark current}
During ground calibration, the detector was cooled down to -31 \textcelsius.
At this temperature, the dark signal was acquired and analyzed to estimate the pixel dark current and its spatial variability also called dark-signal non-uniformity.

\begin{figure}
\centering
 \includegraphics[width=\linewidth]{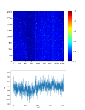}

  \caption{Dark current estimation during in-flight calibration campaigns (2021/03/25). 
  Top figure: Dark current image showing the per-pixel dark current estimated using 
  50-frame onboard average of 10 second integration time images. 
  The color table indicates the dark current value expressed in $e.s^{-1}$.
  Bottom figure: Per-column average dark current corresponding to detector rows.}

  \label{fig:hrieuv_dc_analysis}
\end{figure}

Commissioning in-flight acquisitions allowed to confirm this assessment and complete it with estimation of dark current  as a function temperature during detector cooling.
At operational temperature (-50 \textcelsius) a dark current inferior to 0.05 $e.s^{-1}.pix^{-1}$ was measured (see Figure \ref{fig:hrieuv_dc_analysis}).

Similarly, the electronic gain is confirmed using the dark current at +20 \textcelsius.

%\david{Should we not provide a numerical value for the dark current and plot showing where it comes from? Maybe it comes later?}
% TODO dark current figures

\subsubsection{Detector spectral response}

The detector spectral response was measured at PTB in January 2017. 
Each FM detector  spectral response was measured between 16 nm and 30.4 nm. 
At each wavelength, a total of 10 images was acquired in order to estimate the detector efficiency and effective gain defined hereafter. 
In order to protect the sensor, the EUV beam was localized in the bottom right side of the image area. 
The detector efficiency $D$ is defined as the ratio between the number of photoelectrons and the number of incident photon, measured in $e^-.ph^{-1}$, as follows:

$$
I_{DN} = D \times (I_{ph}) + I_o + \epsilon
$$
where $I$ is the detected pixel signal in DN, $I_{DC}$ is the dark signal,  $I_o$ is the electronic offset,  and $\epsilon$ is random readout signal. 
The mean-variance relation is used to measure the effective gain that relates the temporal variance to the  temporal mean signal. 

\begin{figure}
\centering
 \includegraphics[width=\linewidth]{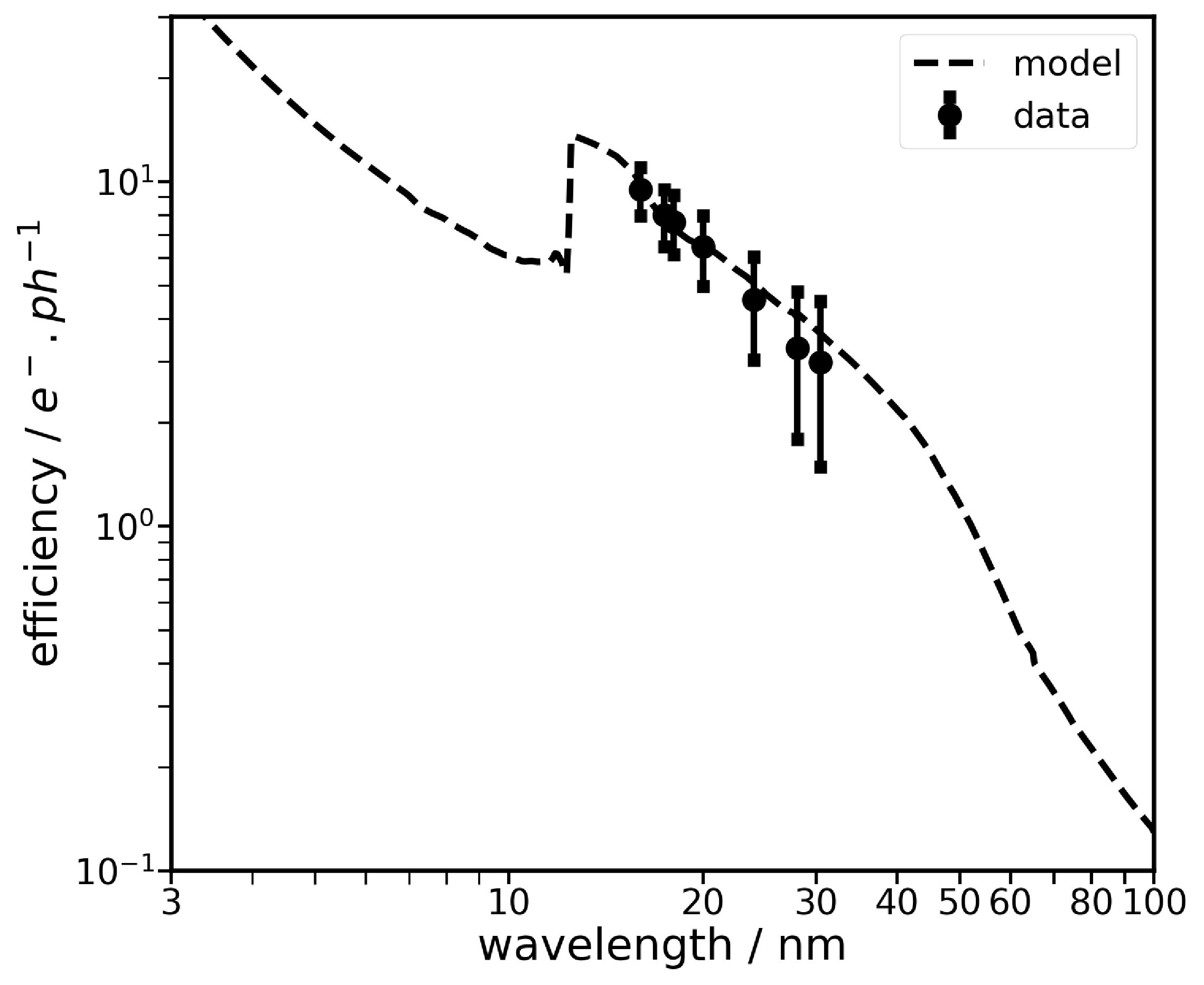}

  \caption{Detector efficiency $D$ of the $\textrm{HRI}_\textrm{EUV}$ detector, with measured data overplotted on the simulated data using detector efficiency model.}

  \label{fig:hrieuv_detector_efficiency}
\end{figure}

The detector gain is here defined as the factor that relates the temporal average of the signal to its variance. The signal gain $g$ relates the signal variance $\sigma_{DN}^2$ to the signal $\mu$ through the ``mean-variance'' law:
$$
\sigma_{DN}^2 = g \times (\mu_{DN} - \mu_{DN,o}) + \sigma_\epsilon^2
$$
where $g$ is the effective gain, $\sigma_{DN}^2$ is the temporal variance of the detected signal, $\mu_{DN}$ its mean, $\mu_{DN,o}$ is the mean offset, and $\sigma_\epsilon^2$ is the readout noise variance.

\begin{figure}
\centering
 \includegraphics[width=\linewidth]{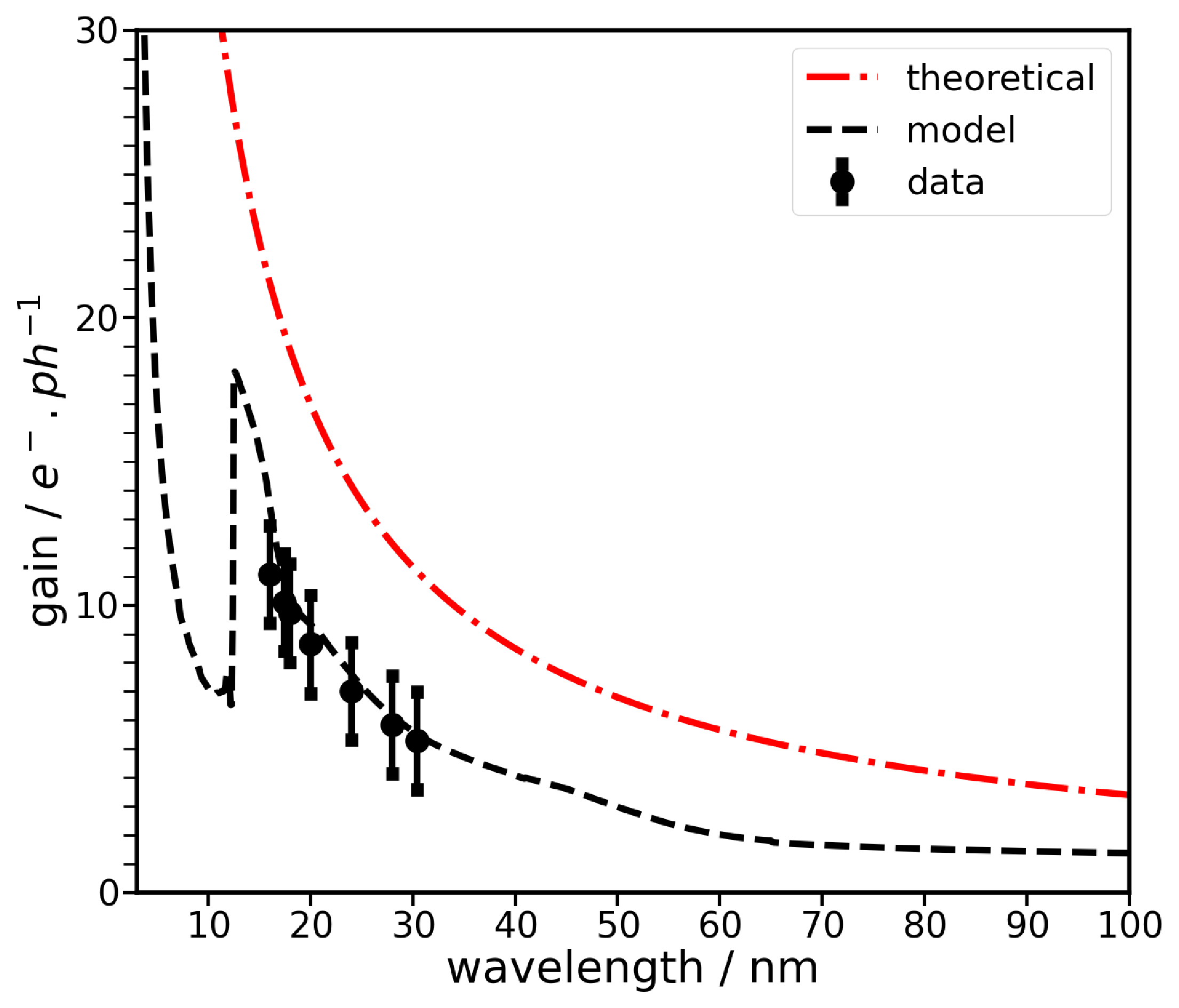}

  \caption{Detector pixel photodiode EUV photon gain of the $\textrm{HRI}_\textrm{EUV}$ detector, with measured data overplotted on the simulated data using detector efficiency model.}

  \label{fig:hrieuv_detector_gain}
\end{figure}

The gain was measured during the detector characterization campaign, at the same wavelength as the efficiency and using the mean-variance relationship. 

$$
\sigma_{e^-}^2 = g(\lambda) \times (\mu_{e^-} - \mu_{e^-,o}) + \sigma_{\epsilon,e^-}^2
$$

where $g(\lambda)$ is the effective gain from EUV photon to electrons and other quantities are expressed in electrons.  
Figure \ref{fig:hrieuv_detector_gain} shows the value of the gain as a function of wavelength, measured and compared to our semi-empirical model-based simulated data. 
This gain is systematically lower than the theoretical value given by the quantum yield predicted value of a fully depleted detectors, in which nearly all electrons can be assumed to be collected in the pixel photodiode.

\subsubsection{Non-linearity}

\begin{figure}
\centering
 \includegraphics[width=\linewidth]{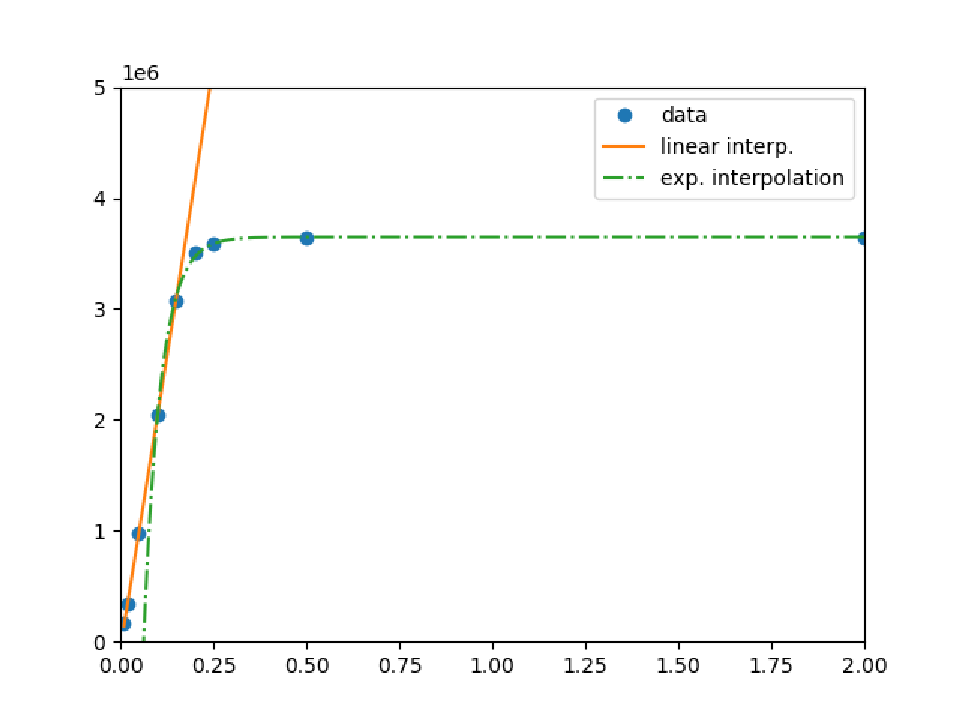}

  \caption{Example of non-linearity estimation of a typical low-gain pixel exposed with blue 470 nm light (locally averaged using binning for statistical reasons). The low-gain signal is measured in DN, here shown as a function of integration time, and modelled by an exponential saturation and compared to the linear regression in the linear range of the sensor. }

  \label{fig:detector-nonlinearity}
\end{figure}

Using in-flight acquired data, based on the visible (blue) onboard calibration LED. 
The non-linearity close to pixel well saturation is assessed by acquiring set of 10 images at multiple increasing integration times. The non-linearity is expressed in low gain DN unit, corresponding resp. to the 1$\%$, 2 $\%$, 5 $\%$ and 10 $\%$ deviation from linear behaviour (see Fig. \ref{fig:detector-nonlinearity}).

The linearity part corresponds to the light flux emitted by the LED at that pixel. 
The ``saturation'' value corresponds to the pixel full well saturation.

\begin{figure}
\centering
 \includegraphics[width=\linewidth]{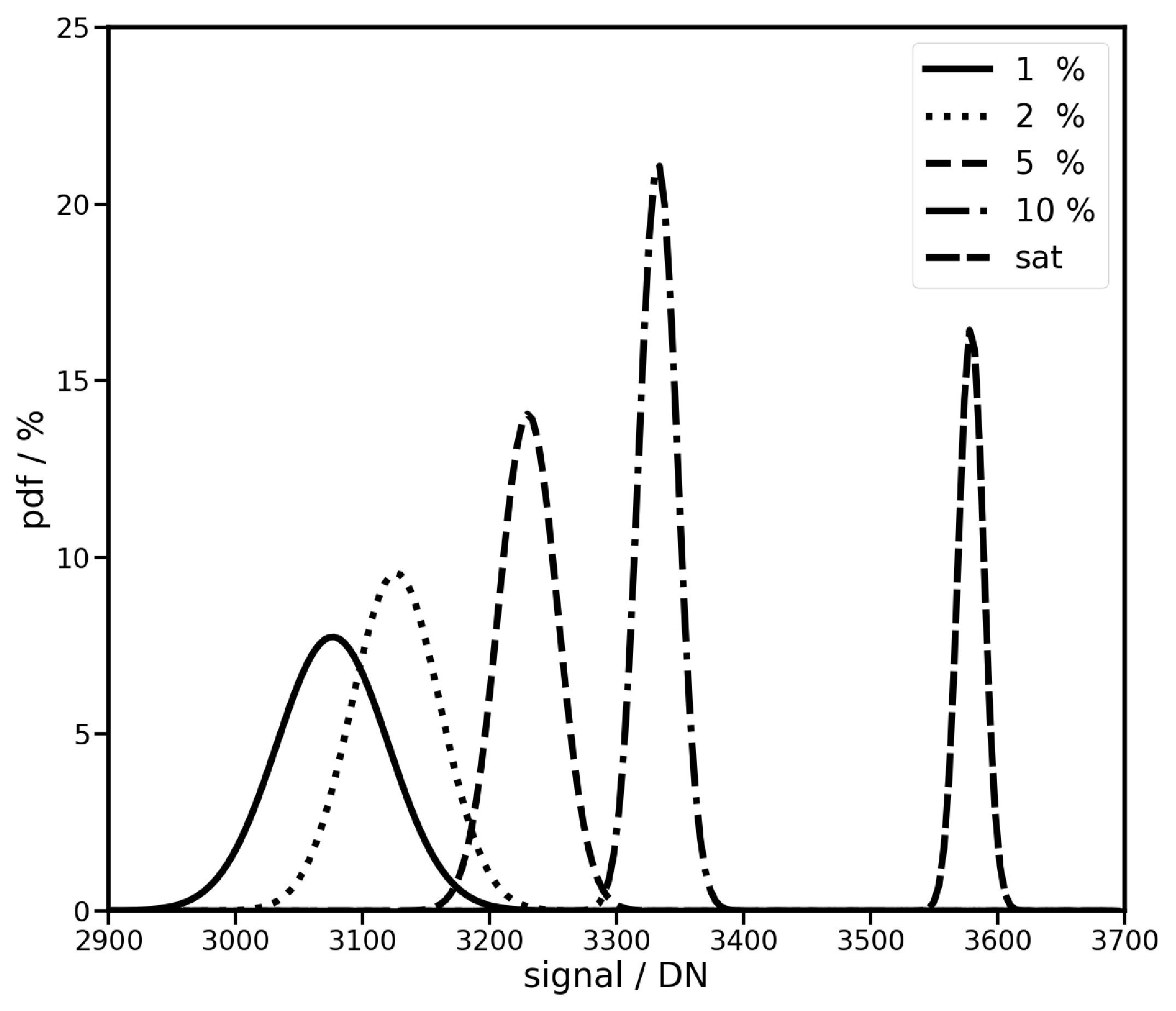}

  \caption{Histogram of pixels for several non-linearity levels. The full well saturation is reached at around 3500 DN, which corresponds to a saturation full well capacity of 131 ke.}

  \label{fig:detector-nonlinearity-hist}
\end{figure}

The non-linearity was assessed on rebinned images, and in a region-of-interest above and close to the stitching line  (in the detector reference frame).

\begin{table}
\caption{\label{tab:t_nonlinearity_val} Detector non-linearity values.}
\centering
\begin{tabular}{lc}
Non-linearity threshold \vline \vline &  Signal in DN (electrons) \\
\hline\hline
1$\%$ &  3058 ([113 ke)\\
2$\%$ &   3106 (115 ke)\\
5$\%$ &  3210 (119 ke) \\
10$\%$ &   3314 (123 ke) \\ 
saturation & 3558 (131 ke)  \\
\hline
\end{tabular}

\end{table}

The non-linearity is clearly reached before the ADC saturation of the low-gain channel (see Table \ref{tab:t_nonlinearity_val}).

\subsubsection{Spatial noise and flat-field}

The spatial noise is characterized by the offset FPN, the dark-signal non uniformity (DSNU) and the photo-response non-uniformity (PRNU).

The APS detector shows a low-signal non-linearity, in particular  at very low integration times.
The detector linearity in the high-gain channel is nearly perfect, while in combined images, at high intensity , the signal becomes non-linear near saturation. 
The non-linearity is estimated using the visible LED in the low gain and is shown in Figure. 

The main cause for the detector full well saturation non-linearity resides in the saturation of the pixel low-gain sense node (or floating diffusion capacitance). Other non-linearities (ADC, source follower) are estimated in 
%TODO include the ADC non-linearities

We use the nominal calibration LED to estimate these parameters despite its disadvantages (non-uniformity, stabilization time).

\begin{figure}
\centering
 \includegraphics[width=\linewidth]{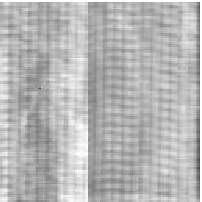}
\vspace{30pt}
  \caption{Detector flat field derived from onboard calibration LEDs. 
  The detector stitching line is visible as an column in the image due to the image rotation.} 

  \label{fig:hrieuv_flatfield}
\end{figure}

The detector flat-field could not be measured using EUV light. 
Instead, it was derived from onboard calibration LED measurements that were averaged and processed on-ground (see Figure \ref{fig:hrieuv_flatfield}). 

\subsection{Spectral response}
The multiplication of component performance functions in Eq (\ref{eq:spectral_response}) gives the model of the instrument performance  as shown in Figure \ref{fig:detector-nonlinearity}. 
%TODO Add spectral lines for comparisons

\begin{figure}
\centering
 \includegraphics[width=\linewidth]{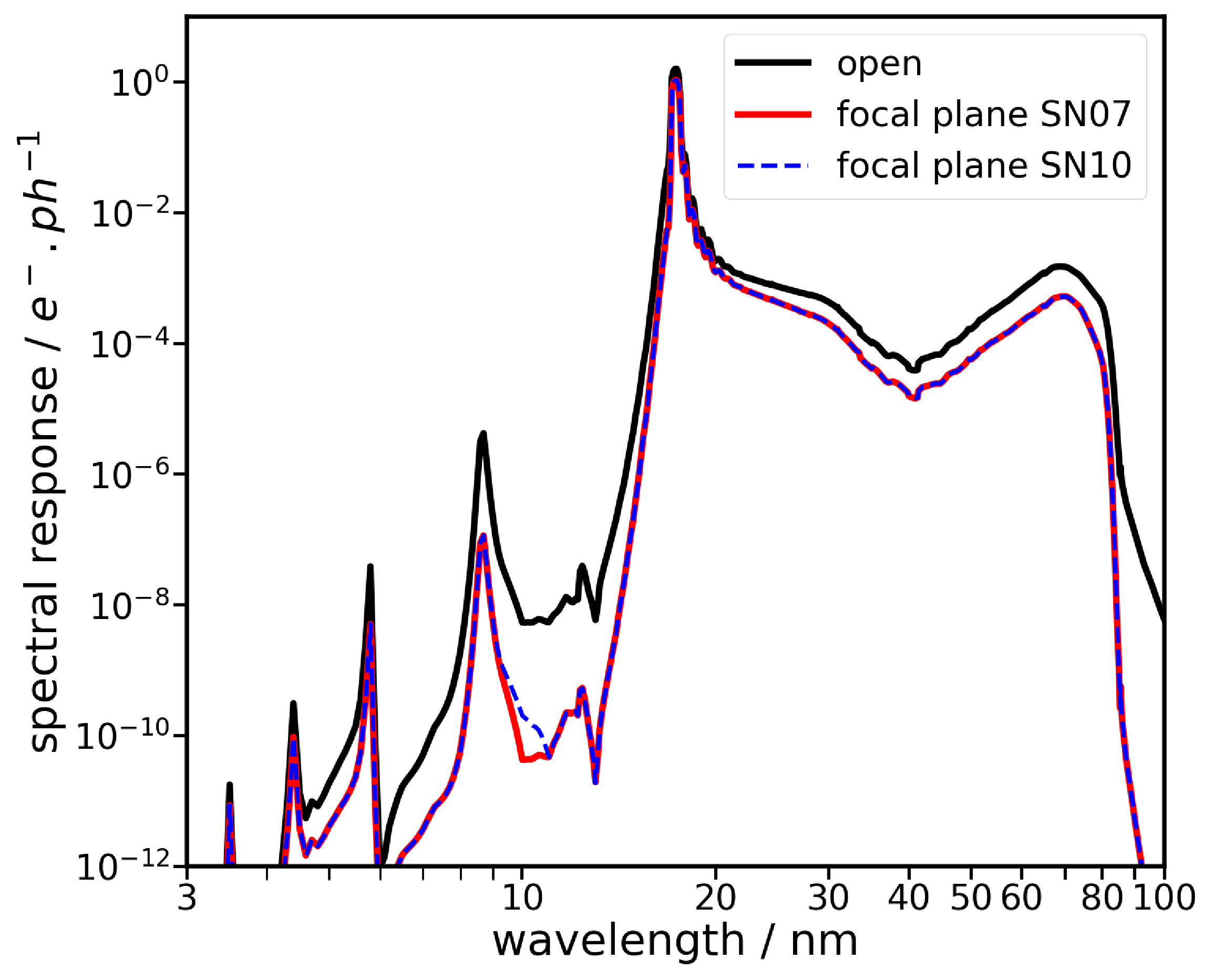}

  \caption{Spectral response simulation using the semi-empirical model of the $\textrm{HRI}_\textrm{EUV}$ instrument in the 1 nm-100 nm wavelength range, in electron per incident photon unit.}

  \label{fig:spectral_response_1}
\end{figure}

%TODO include the temperature response using Chianti

\subsection{Camera characterization and detector setting optimization}

In order to allow for the onboard reconstruction of high- and low-gain images, the detector offset must be low enough so that the matching is possible between both channels and gain combination can be performed onboard in the Common Electronics Box (CEB). 
In the reconstruction process the threshold-selected low-gain pixels scaled to the high-gain channel intensity scale. 
The original detector settings used during ground calibration was updated in-flight during commissioning in order to optimize the low and high gain offsets. 
As a result of this modification, the dark frames used for calibration have significantly lower values, with offsets of few DNs. 

Due to the late The $\textrm{HRI}_\textrm{EUV}$ detector change, the EUV flat-field of the detector is not part. 
But due to  similar photon attenuation depth, the flat-field obtained using the calibration LEDs (see \ref{sec:calibration_LED}) correlates very well with the EUV flat-field which was confirmed by measurements of EUI/FSI data which has the same CMOS APS back-side illuminated technology.

\section{End-to-end calibration}
\label{sec:ground_calibration}

The end-to-end EUV calibration campaign was performed at PTB on April 25, 2017 at the Metrology Light Source of PTB, Berlin (Germany). 
The calibration was performed using the MLS U125  beamline\citep{Gottwald2019}.
The EUI optical structure with the three telescopes were installed in the PTB vacuum chamber  shown on
Figure and  mounted on a dedicated support bracket fixed in the chamberon the translation and rotation stage of the chamber, allowing to scan the channels FOV.

%TODO : uncertainty budget
% 
% \begin{table*}
% \caption{\label{t3} Relative uncertainty budget for a coverage factor k = 2 (2-s uncertainty).}
% \centering
% \begin{tabular}{lccc}
% Component & Parameter (\%) \vline \vline & value & uncertainty \\
% \hline\hline
% Aperture
% Error entrance filter (\%) \vline \vline &  diameter / thickness / transmission &  5 & \\
%  & \\
%  & \\
% & \\
% \hline\hline
% &\\
% &\\
%  & \\
%  & \\
% \hline
% \end{tabular}
% 
% \end{table*}

%Resulting total photon flux 3.5 2.2

\begin{figure}
 \includegraphics[width=\linewidth]{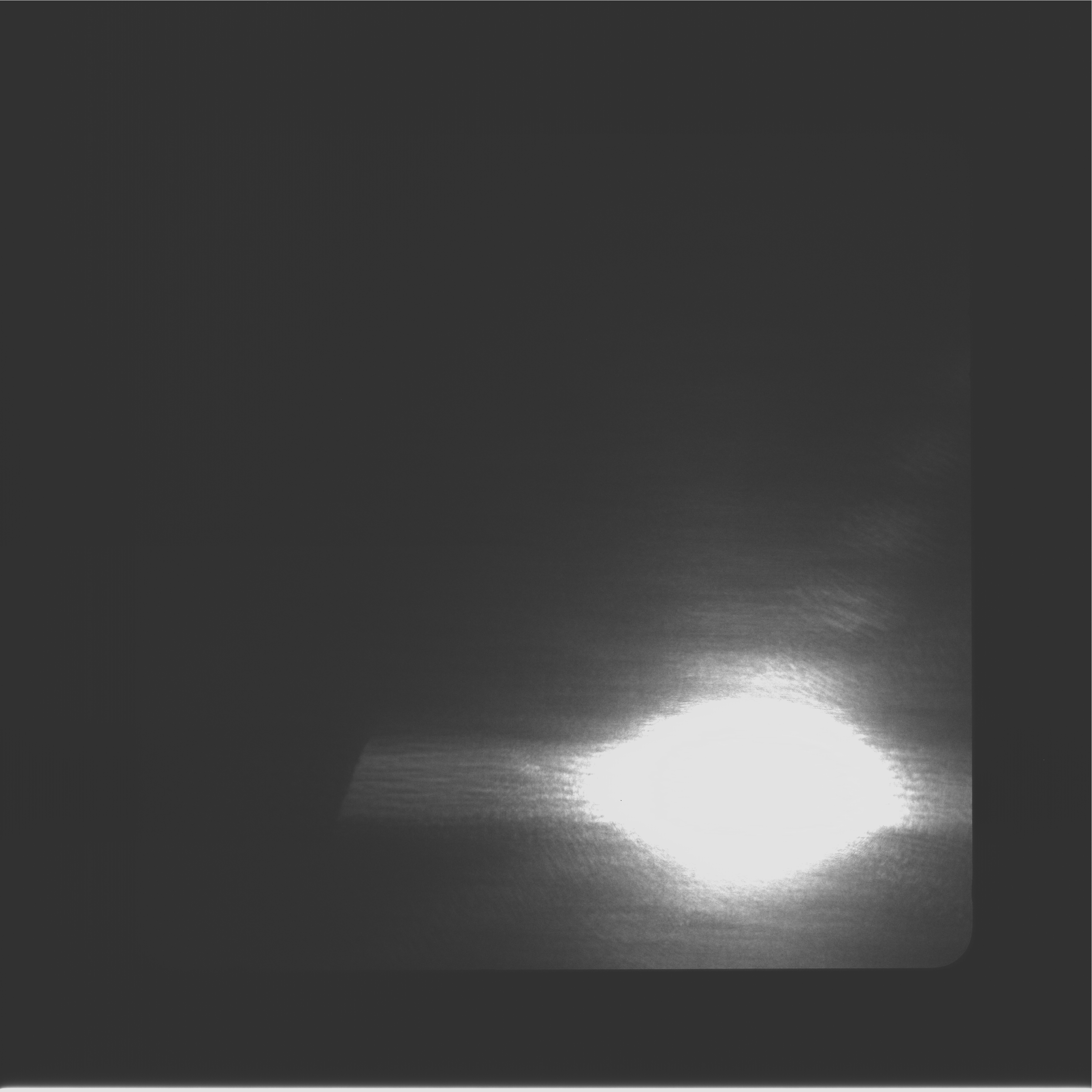}

  \caption{$\textrm{HRI}_\textrm{EUV}$ first image acquired during ground calibration. Bottom right, the occultation of the EUV beam in the detector plane is visible.}

  \label{fig:hrieuv_first_light}
\end{figure}

After beam centering and first image acquisitions (see Fig. \ref{fig:hrieuv_first_light}), it was decided to lower the beam flux in order to protect the detector from any EUV degradation before calibration.
In the synchrotron ondulator configuration, the beam divergence is lower than 2 microrad \citep{Gottwald2019}. 

A spatial scanning procedure was applied by moving the instrument according to $x$ and $y$ axis which also allowed to check the aperture size along the two dimensions.  
% Field-of-view scanning 

After the first calibration light image was acquired, 
a short manual procedure was performed to align the EUV beam with the detector centre, we immediately decreased the light flux in order to protect the detector sensitive surface.
%TODO entrance filter geometry

\subsection{Spectral response}
After the vacuum pumping of the tank where the EUI instrument was accommodated, the detector temperature was stabilized to -35 \textcelsius. 

After a set of attempts to orient the with the $\textrm{HRI}_\textrm{EUV}$ telescope  so that the EUV beam is centered in the image, the first light image was acquired was acquired with synchrotron light beam in nominal conditions and high flux with the detector in high gain, original settings. In order to find the beam easily, the EUV beam was in high flux configuration, up to 1E9 ph.s$^{-1}$. After this first centering step, the beam flux was lowered to 1E6 ph.s$^{-1}$ in order to protect the sensor from early degradation. 

For the first part of the measurements, the filter wheel was set in open position. 

This allowed to scan the entrance aperture and confirm its dimension of 47 mm.

Images are acquired in single-exposure mode, and using the FM instrument electronics.

The end-to-end instrument response was measured by acquiring 10 images of the EUV beam at each wavelength between 16 and 18.5 nm with a 0.1 nm sampling between 17 and 17.6 nm. 

During end-to-end calibration, the 17.4 nm high-flux beam was off-pointed at (resp.) 0.5, 1, and 1.5 degree along x-axis to measure stray light. No significant stray light could be measured.

\subsection{Dark noise}

During commissioning, the sensor was first heated using the decontamination heater in order to protect the detector surface from possible contamination. On April 14, 2020, the cooling started and during which the detector reached a temperature of -50 \textcelsius.  This enabled the estimation of the dark current as a function of temperature. 

\subsection{EUV and visible flat field}

As opposed to the FSI telescope, the EUV flat-field of the $\textrm{HRI}_\textrm{EUV}$ image sensor is not available due to the breaking of the sensor and its replacement by the spare model. 
Nevertheless, in-flight visible light measurements thanks to the onboard calibration LED could be used to estimate visible flat-field. Using FSI EUV flat-field and equivalent 470nm emitting blue LED, it confirmed the high correlation between EUV and visible flat fields. This is due to the fact that penetration depth of EUV photons at near 17.4 nm is comparable to 470 nm photons.

\begin{figure}
 \includegraphics[width=\linewidth]{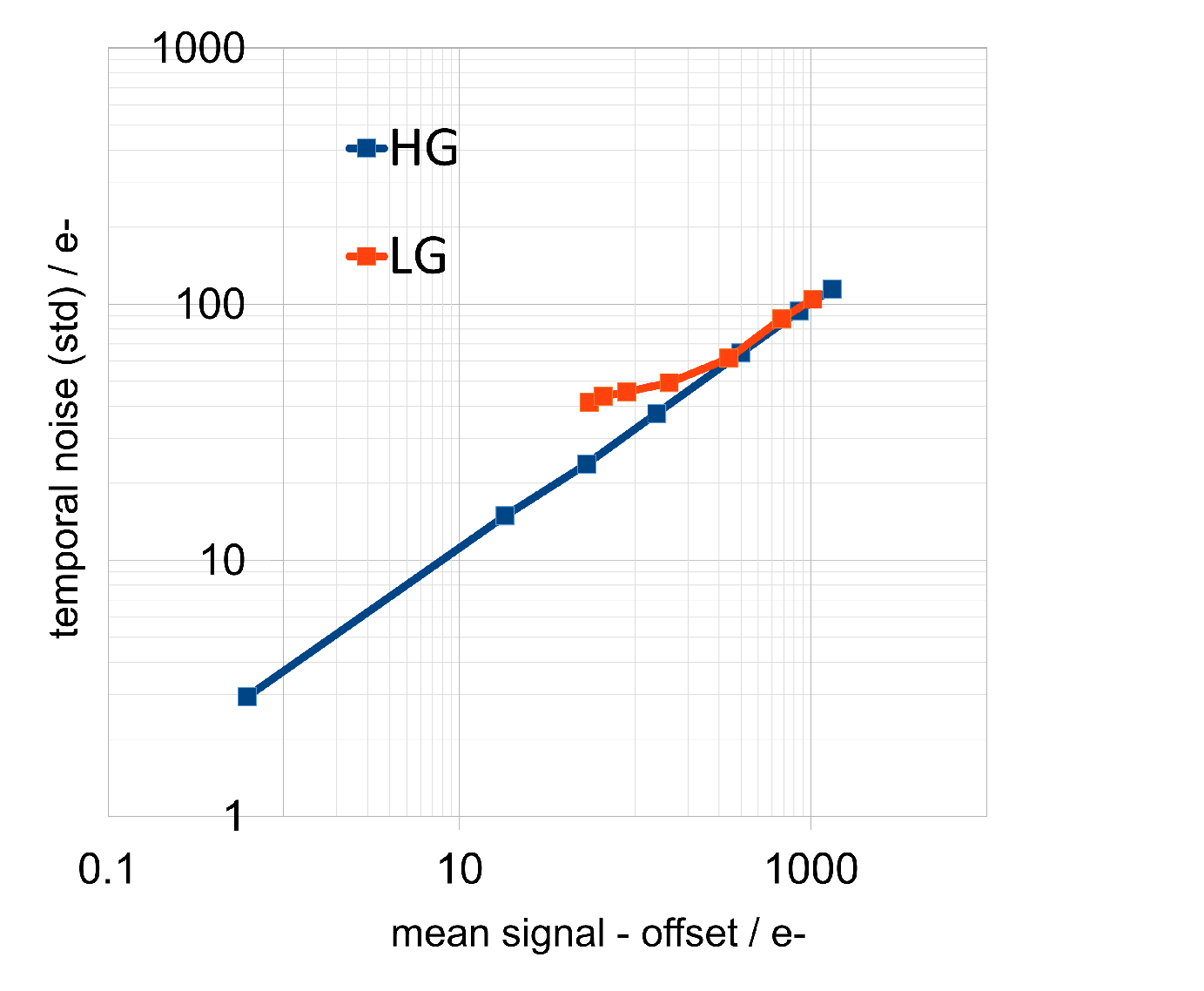}

  \caption{PTC curves showing the matching of the high- and low-gain pixel values obtained with EUV light at 17.4 nm. The high-gain lowest detected signal has a signal sigma of 3 e- rms, which corresponds to its expected readout noise, at the detection limit. Due to the protection measure on the detector, the EUV flux was kept low so the saturation was not reached and no full well saturation is visible.}
  \label{fig:hrieuv_gain_matching}
\end{figure}

\begin{figure}
\centering
 \includegraphics[width=6.5cm,angle=90]{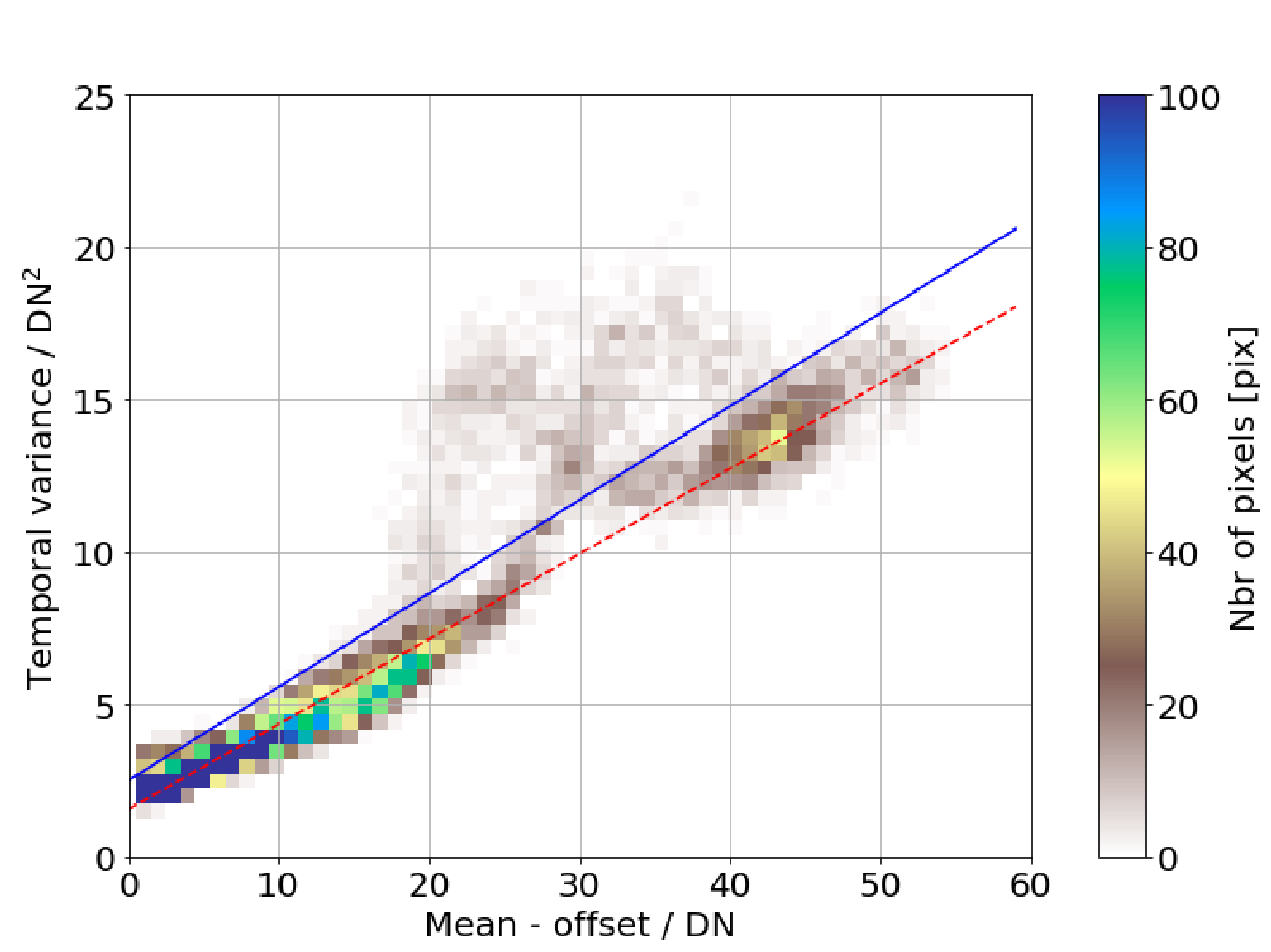}

  \caption{Single integration time  image PTC estimation of the low gain with EUV beam light. The robust red linear fit give a photon to DN low gain value of 0.279 $DN.ph^{-1}$. Pixels located to the pattern above the linear regression line undergo the ``ripple effect''.}

  \label{fig:EUV_low_gain}
\end{figure}

The EUV high and low gain were measured by computing the per-pixel temporal mean and variance of set of 10 images per integration time. The fitting of the variance as a function of mean signal, after dark signal correction,  for each wavelength, gives the effective gain that is shown in Figure . 
Another method to estimate the spatial average of the EUV gain is to fit the variance of, taking advantage of the EUV beam non-uniformity, at a single wavelength (Figure ), for high- and low-gain (Fig. \ref{fig:EUV_low_gain}).

\begin{figure}
\centering
 \includegraphics[width=8cm]{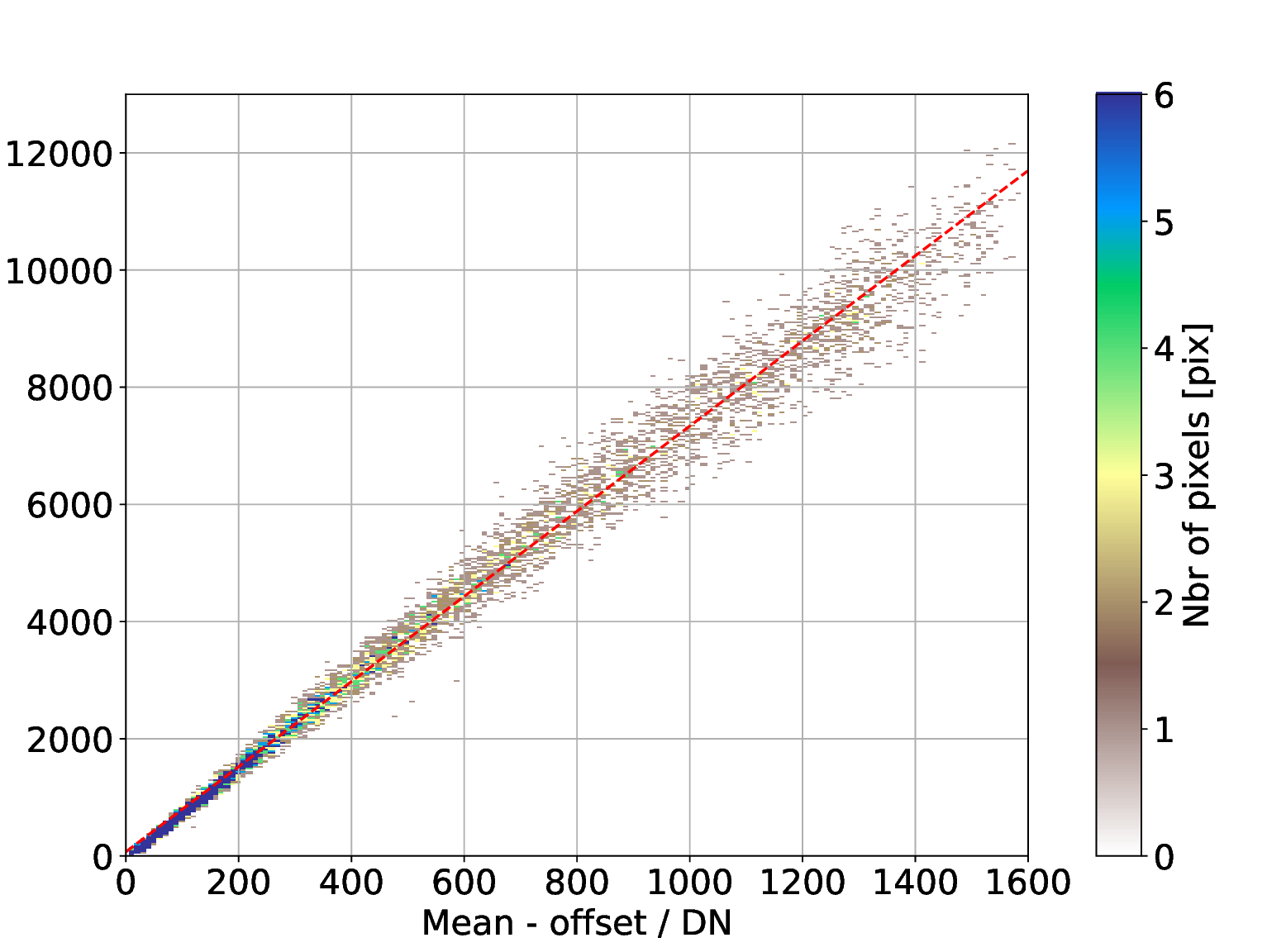}
\vspace{30pt}
  \caption{Single integration time image PTC estimation of the high  gain with EUV beam light. The robust red linear fit give a photon to DN low gain value of 7.29 $DN.ph^{-1}$.  This corresponds to a gain ratio of $\approx$ 26.}

  \label{fig:EUV_high_gain_hist}
\end{figure}

\begin{figure}
\centering
 \includegraphics[width=\linewidth]{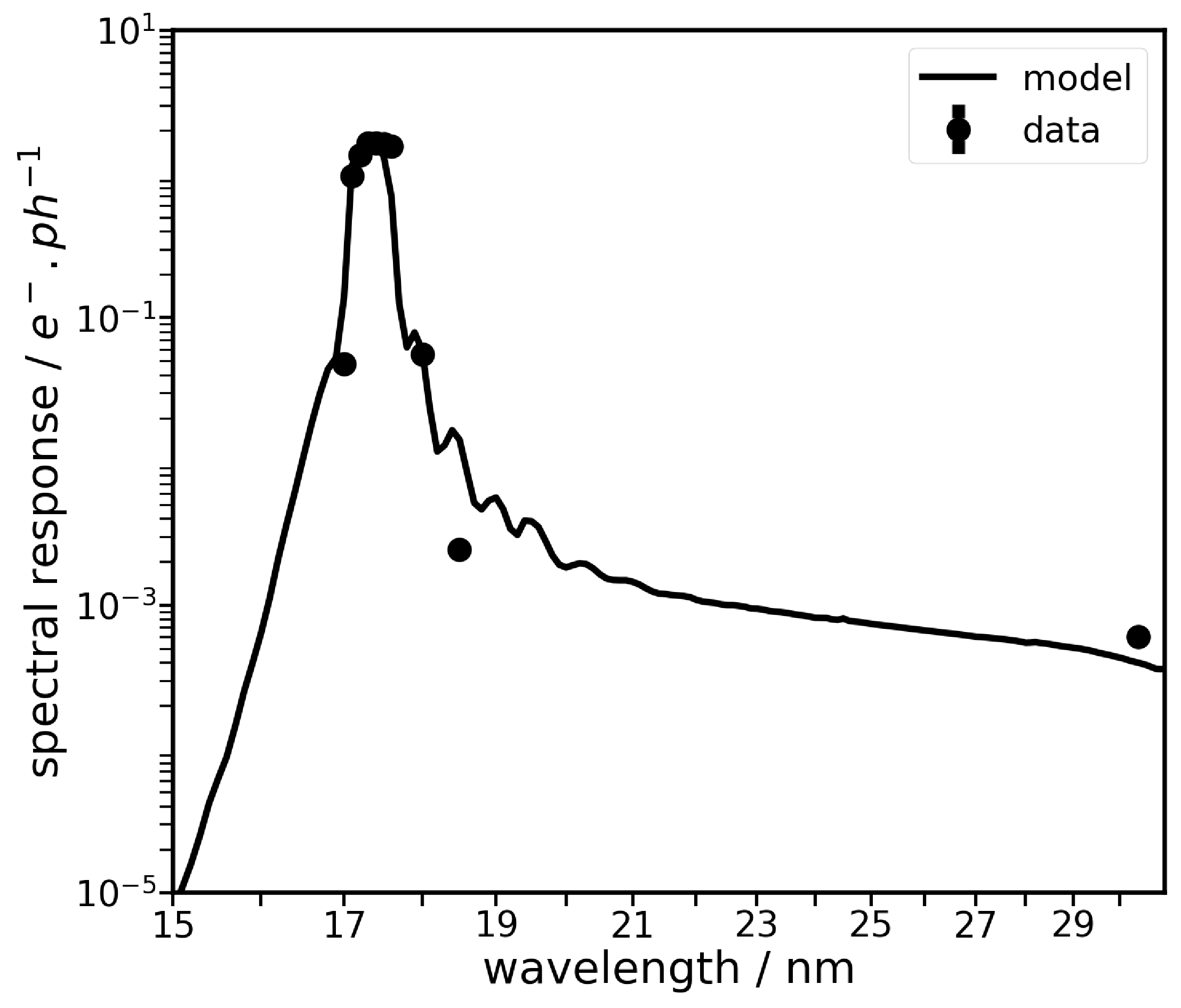}

  \caption{Spectral response of the $\textrm{HRI}_\textrm{EUV}$ instrument measured during ground calibration, including the 30.4 nm wavelength measurement, with the filter wheel in open position.}

  \label{fig:spectral_response_2}
\end{figure}

\begin{figure}
\centering
 \includegraphics[width=\linewidth]{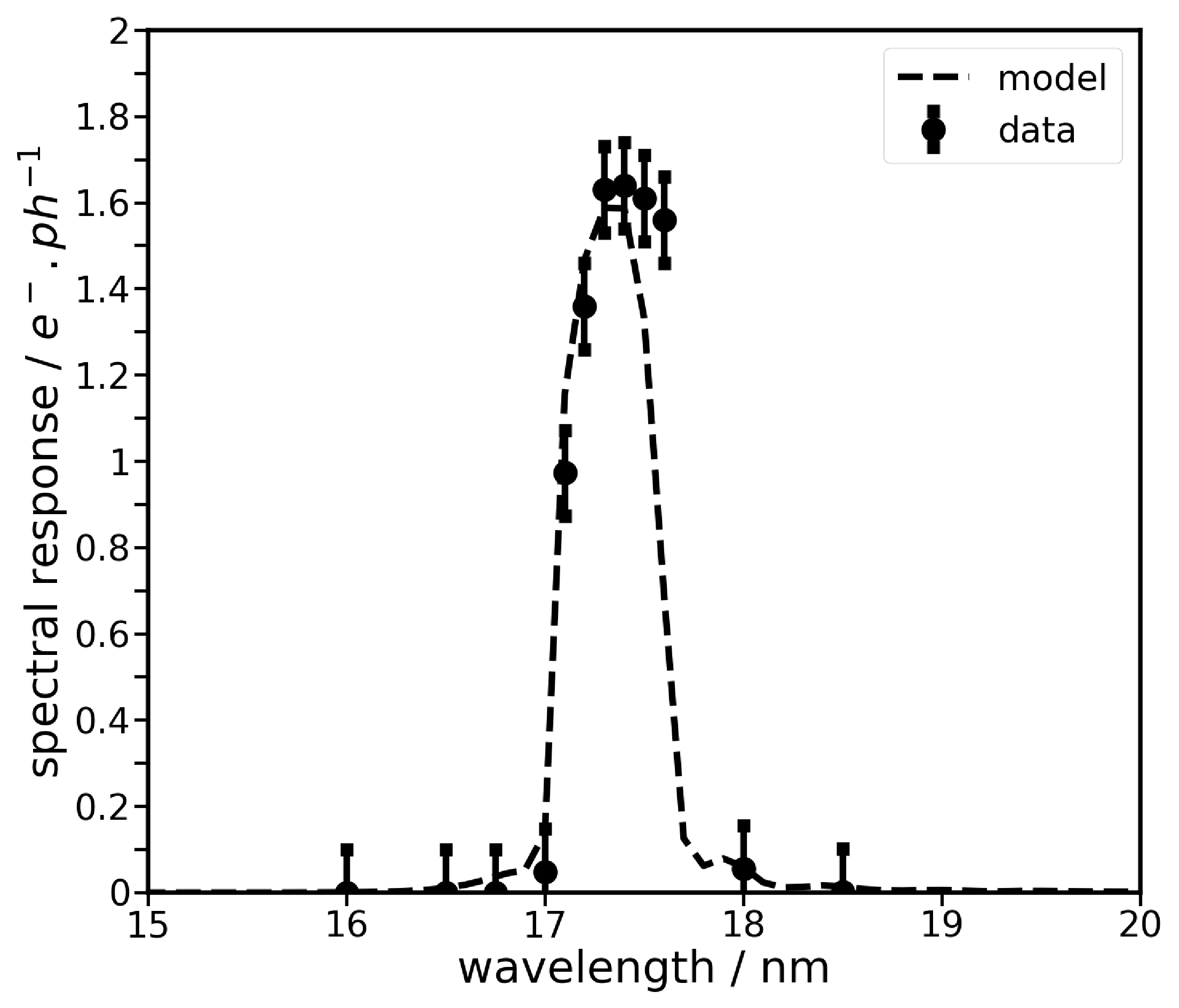}

  \caption{Spectral response of the $\textrm{HRI}_\textrm{EUV}$ instrument measured during ground calibration.}

  \label{fig:spectral_response_3}
\end{figure}

\begin{table*}
\caption{\label{tab:perf_summary} $\textrm{HRI}_\textrm{EUV}$ telescope measured performance summary. 
Values in DN correspond to 12-bit scale of the corresponding low-or high gain of the detector.}
\centering
\begin{tabular}{lcc}
Parameter &  Measured value  & Error  \\
\hline\hline
Peak wavelength & 17.4nm & \\
Spectral response (17.4nm) &> 1.6 $e.ph^{-1}$& \\
platescale & 0.492 arcsec & \\
telescope focal length & 4187 mm & \\
Resolution (pixel) & 0. 49 arcsec & \\
Read noise (HG) & 2.8 e () rms & \\
Dark current (pix)& < 0.05 $e^.s^{-1}$ @ -50 \textcelsius  & \\
Electronic gain, HG (mean) & 0.64 $DN.e^{-1}$ (1.6 $e.DN^{-1}$) & 10\%\\
Electronic gain, LG (mean) & 0.027 $DN.e^{-1}$ & 10\%\\
Non-linearity sat (LG)& >130 ke (>3558 DN) & \\
EUV HG Gain (mean), 17.4 nm &  7.4  $DN.e^{-1}$ (model) & 10 $\%$ \\
EUV HG Gain (mean), 17.4 nm &  6.9  $DN.e^{-1}$ (method 1) & \\
EUV HG Gain (mean), 17.4 nm &  7.3  $DN.e^{-1}$ (method 2) & \\
PRNU  & < 2 $\%$ & \\
\hline
\end{tabular}

\end{table*}

% \begin{quote}
% \begin{itemize}
% \item Full well >100 000 electrons
% \item Dark current <1 electrons $s^-1$ at -18 \textcelsius
% \item Linearity  XX 
% \item Readout rate (max.)  pixels s-1
% \item Power < 1 Watt
% \item Amplifier gain 12 e/DN, amplifier A (default)
% \item Read noise XX electrons
% \item Resolution 12 bits
% \item Detector (column) offset DN (1 × 1 binning), 4-column pattern
% \end{itemize}
% \end{quote}

\subsection{Detector artifacts}
The HRI-EUV detector has deficient pixels and artefacts. 
A ``ripple'' effect visible in the low-gain pixels consists in signal ``bump'' and drop at a certain distance of a high signal with which it seems to correlate well. Since this was discovered after focal-plane assembly and just before end-to-end calibration, it could not be well characterized on-ground. 

A small area of the detector  contains deficient pixels (``jelly baby'') that are highly non-linear. These pixels will be corrected by the on-ground calibration routine (euiprep). 
On the edge of the detector (left band in the detector frame), the signal is falling rapidly to below zero values. 

A few bad (hot) pixels were identified on-ground.

\section{Instrument spectral response}
\label{sec:spectral_response}

\subsection{In-flight EUV efficiency}
%TODO inflight-data PTC
In order to measure the EUV efficiency, the gain is estimated from sequence of EUV images.
 It should be noted that the sequences used in this estimation suffer both solar signal variability and instrument jitter.
 Nevertheless, this method is used to confirm in-flight the pixel gain with EUV photons.
The obtained gained is inherently dependent on the solar variability of the structure that lies within the pixel area. 

This limits the precision of the values that are obtained.
%  \begin{figure}
%    \includegraphics[width=\linewidth]{./figs/hrieuv_gain1.eps}
% 
%   \caption{$\textrm{HRI}_\textrm{EUV}$ EUV gains and response}
% 
%   \label{fig:hrieuv_EUV_gain1}
% \end{figure}

\subsubsection{Onboard image processing}

The $\textrm{HRI}_\textrm{EUV}$ images are Poisson-recoded and can be either lossless or lossy compressed onboard. 
Only few uncompressed data have acquired during calibration and commissioning due to low telemetry.

\subsection{Ground processing}
Telemetry packets are processed on-ground and depacketized to Level 0 FITS image format. 
Elementary corrections are implemented and metadata including necessary housekeeping information are incorporated to generate the L1-level FITS files. 
Calibrated data are generated using the euiprep Python routine.

The end-to-end telescope response is shown in Figs \ref{fig:spectral_response_2} and\ref{fig:spectral_response_3}.

\section{Conclusion}
\label{sec:conclusion}

The initial preflight radiometric calibration of the $\textrm{HRI}_\textrm{EUV}$ has confirmed the very high BOL sensitivity of the telescope (>1.6 electrons per EUV photons at peak wavelength) when compared to the low dark current at operating temperature (<-50 \textcelsius) and the low high-gain read out noise (2.8 electrons rms). 

For in-flight operations all necessary products are now available including the dark frames with new detector settings and a flat field derived from LED measurements and applied on-ground by the L1-to-L2 euiprep calibration procedure. 

\begin{acknowledgements}
EUI was conceived by a multi-national consortium and proposed in 2008 under the scientific lead of Royal Observatory of Belgium (ROB) and the engineering lead of Centre Spatial de Li\`ege (CSL). The ROB and CSL institutes have received project-specific funding from the Belgian Federal Science Policy Office (BELPSO).
\end{acknowledgements}

\bibliographystyle{aa} % style aa.bst
\bibliography{./bibtex/biblio} % your references Yourfile.bib

\end{document}